\documentclass[useAMS,usenatbib]{mn2e}
\usepackage{graphicx}
\usepackage{setspace}
\usepackage{natbib}
\usepackage{color}
\usepackage{amsmath,amssymb}
\usepackage{times}
\usepackage{aas_macros}

\newcommand\umg{u\!-\!g}
\newcommand\gmr{g\!-\!r}
\newcommand\rb{r_{\rm b}}
\newcommand\alphain{\alpha_{\rm in}}
\newcommand\alphao{\alpha_{\rm out}}
\newcommand\rellip{r_q}
\newcommand\reff{r_{\rm eff}}

\title[The Milky Way Stellar Halo]{The Milky Way stellar halo out to
  40 kpc : Squashed, broken but smooth} \author[A. J. Deason,
V. Belokurov and
N. W. Evans]{A.J. Deason$^{1}$\thanks{E-mail:ajd75,vasily,nwe@ast.cam.ac.uk},
  V. Belokurov$^{1}$ and N. W. Evans$^{1}$\\ $^{1}$Institute of
  Astronomy, Madingley Rd, Cambridge, CB3 0HA}

\begin{document}


\date{July 2011}

\pagerange{\pageref{firstpage}--\pageref{lastpage}} \pubyear{2011}

\maketitle

\label{firstpage}

\begin{abstract} 
  We introduce a new maximum likelihood method to model the density
  profile of Blue Horizontal Branch and Blue Straggler stars and apply
  it to the Sloan Digital Sky Survey Data Release 8 (DR8) photometric
  catalogue. There are a large number ($\sim$ 20,000) of these tracers
  available over an impressive $14,000$ deg$^2$ in both Northern and
  Southern Galactic hemispheres, and they provide a robust measurement
  of the shape of the Milky Way stellar halo.  After masking out stars
  in the vicinity of the Virgo Overdensity and the Sagittarius stream,
  the data are consistent with a smooth, oblate stellar halo with a
  density that follows a broken power-law. The best fitting model has
  an inner slope $\alphain \sim 2.3$ and an outer slope $\alphao \sim
  4.6$, together with a break radius occurring at $\sim 27$ kpc and a
  constant halo flattening (that is, ratio of minor axis to major axis)
  of $q \sim 0.6$.  Although a broken power-law describes the density
  fall-off most adequately, it is also well fit by an Einasto
  profile. There is no strong evidence for variations in flattening
  with radius, or for triaxiality of the stellar halo. 
\end{abstract}


\section{Introduction}

The time required for stars in the stellar halo to exchange their
energy and angular momentum is very long compared to the age of the
Galaxy.  Therefore, such stars preserve memories of their initial
conditions, and so the structure of the stellar halo is intimately
linked to the formation mechanism of the Galaxy itself.  This
fundamental insight was already noted by \cite{eggen62}. It is the reason why the stellar halo has attracted such
interest despite containing only a small fraction of the total stellar
mass of the Galaxy. Stars diffuse more quickly in configuration space,
as opposed to energy and angular momentum space. So, the spatial
structure of the stellar halo may be smooth, even though it is built
up from merging and accretion.

The simplest way of studying the stellar halo is through
starcounts. Typically, RR Lyrae or blue horizontal branch stars (BHBs)
are used as tracers, as they are relatively bright ($M_g \sim 0.5-0.7$, e.g. \citealt{sirko04}) and can be detected
at radii out to $\sim$ 100 kpc. The gathering of such data is
painstaking work, and carries the price that sample sizes are often
small.  Such studies are consistent with a stellar halo that is round
in the outskirts (with a minor-to-major axis ratio $q =1$) and more
flattened in the inner parts with $q \sim 0.5$ (e.g.,
\citealt{hartwick87}; \citealt{preston91}). Rather than selecting
typical halo stars, an alternative approach is to model deep star
count data in pencil-beam surveys at intermediate and high galactic
latitudes, allowing for contamination of the starcounts by the thin
and thick disk populations. This was attempted by \cite{robin00}, who
found a best-fit halo density law with flattening $q \sim 0.76$,
together with a power-law fall-off $\alpha$ of $2.4$ (that is, $\rho
\sim ({\rm distance)}^{-\alpha}$). A similar, slightly later, attempt
by \cite{siegel02} using data in seven Kapteyn selected areas
yielded $q \sim 0.6$ and $\alpha \sim 2.75$.

Efforts to detect variations in the flattening with radius have also
been undertaken.  \cite{preston91} argued that the flattening changes
from strongly flattened ($q=0.5$) at 1 kpc to almost round at $20$
kpc.  However, work by \cite{sluis98} using a compilation of 340 RR
Lyraes and BHBs found a constant flattening of $q \sim 0.5$ with no
evidence for changes with radius, as well as and a power-law index
$\alpha \sim -3.2$.  The most recent work by \cite{propris10}
utilising 666 BHB stars from the 2dF Quasar Redshift Survey find that
the halo is approximately spherical with a power-law index of
$\alpha=2.5$ out to $\sim 100$ kpc.  Similarly, \cite{sesar11}
studying Main Sequence Turn-Off (MSTO) stars from the
Canada-France-Hawaii Telescope Legacy Survey find that the flattening
is approximately constant at $q \sim 0.7$ out to $35$ kpc.

The Sloan Digital Sky Survey (SDSS) has transformed our knowledge of
the stellar halo. Although it had been suspected that the stellar halo
is criss-crossed with streams and substructures ever since the
discovery of the disrupting Sagittarius (Sgr), the SDSS provided a
memorable picture of the debris in {\it The Field of Streams}
(\citealt{belokurov06}). A wealth of substructure has now been
identified, including the Sagittarius stream, the Virgo Overdensity
and the Hercules-Aquila Cloud (e.g. \citealt{ibata95};
\citealt{belokurov06}; \citealt{juric08}; \citealt{Be07}). This has
been seen as vindication of modern theories of galaxy formation, which
predict that stellar haloes are built up almost exclusively from the
debris of disrupting satellites (e.g. \citealt{bullock05};
\citealt{delucia08}; \citealt{cooper10}). A number of studies have
attempted to model the smooth halo component by avoiding these known
substructures (e.g. \citealt{juric08}). The results are only in very
rough agreement, suggesting that the density profile of the Milky Way
has a power-law slope in the range $2 < \alpha < 4$ and a flattening
varying from $0.4 < q < 0.8$ (e.g. \citealt{yanny00};
\citealt{chen01}; \citealt{newberg06}; \citealt{juric08};
\citealt{sesar11}).

Even though panoramic photometric surveys like SDSS do provide a large
sample of stellar halo tracers over a considerable portion of the sky,
there is no consensus on the flattening and shape of the stellar halo.
MSTO stars are commonly used tracers owing to their large numbers and
the ease by which they can be photometrically identified
(e.g. \citealt{bell08}). The absolute magnitudes of such stars centre
around $M_r \sim 4.5$ but have a wide range of values ($\sigma_{M_r}
\sim 0.9$ mag) which limits the accuracy to which the density profile
can be estimated.  BHB stars are superior
distance estimators ($\sigma_{M_r} \sim 0.15$), but are significantly
scarcer than main sequence stars. Moreover, they suffer from
contamination by blue straggler (BS) stars due to their similar
colours. BHB and BS stars can be distinguished by their Balmer line
profiles (e.g. \citealt{kinman94}; \citealt{yanny00};
\citealt{sirko04}; \citealt{clewley02}), but this requires
spectroscopic information. Whilst spectroscopic samples can cleanly
identify BHB stars, the variety of results on flattening (e.g.,
\citealt{preston91}; \citealt{sluis98}; \citealt{propris10}) and density fall-off (e.g. \citealt{xue08}; \citealt{brown10}) suggests
that the completeness biases are difficult to understand and control.

An independent constraint on the density profile of the stellar halo
is provided by the velocity distribution of the halo stars. A
kinematic analysis by \cite{carollo07} (see also \citealt{carollo10})
suggests that the stellar halo comprises of two components with
different density profiles and metallicities. The authors find that
the density profile becomes shallower beyond $15-20$ kpc. This is in
stark contrast to the studies by \cite{watkins09} and \cite{sesar10},
who find a significantly steeper density profile beyond $\sim 30$ kpc
from the distribution of RR Lyrae stars in SDSS Stripe 82. A caveat to
the interpretation of kinematic studies is that the density
distribution is not measured directly but rather modelled by assuming
a dark matter halo potential.

Therefore, the present state-of-play is distressingly inconclusive and
a further attack on the problem of the shape of the stellar halo is
warranted.  In this study, we introduce a new method to model both BHB
and BS stars based on photometric information alone. We make use of
the SDSS DR8 photometric data release which has now mapped an
impressive $\sim 14,000$ deg$^2$ of sky with both Northern and
Southern coverage. In contrast to previous work, we combine both the
wide sky coverage of the SDSS with the accurate distance estimates
provided by the BHB stars to model the density profile of the stellar
halo out to $\sim 40$ kpc.

The paper is arranged as follows.  In \S2.1, we describe the SDSS DR8
photometric data and our selection criteria for A-type stars. The
remainder of \S2 introduces the probability distribution for BHB and
BS membership based on colour alone and outlines the absolute
magnitude-colour relations for the two populations. In \S4, we
describe our maximum likelihood method to determine the density
profile of the stellar halo and in \S4 we present our
results. Finally, we draw our main conclusions in \S5.


\section{A-type stars in SDSS Data Release 8}

\subsection{Data Release 8 (DR8) Imaging}

The Sloan Digital Sky Survey (SDSS; \citealt{york00}) is an imaging
and spectroscopic survey covering roughly $\sim 1/4$ of the
sky. Imaging data is obtained using a CCD camera (\citealt{gunn98}) on
a 2.5 m telescope (\citealt{gunn06}) at Apache Point Observatory, New
Mexico. Images are obtained simultaneously in five broad optical bands
($ugriz$; \citealt{fukugita96}). The data are processed through
pipelines to measure photometric and astrometric properties
(\citealt{lupton01}; \citealt{smith02}; \citealt{stoughton02};
\citealt{pier03}; \citealt{ivezic04}; \citealt{tucker06}). The SDSS
DR8 release contains all of the imaging data taken by the SDSS imaging
camera and covers over $\sim14,000$ deg$^2$ of sky (\citealt{dr8}). We
select objects classified as stars with clean photometry. The
magnitudes and colours we use in the following sections have been
corrected for extinction following the prescription of
\cite{schlegel98}.

\begin{figure*}
\centering
  \begin{minipage}{\linewidth}
    \centering
    \includegraphics[width=15cm,height=5cm]{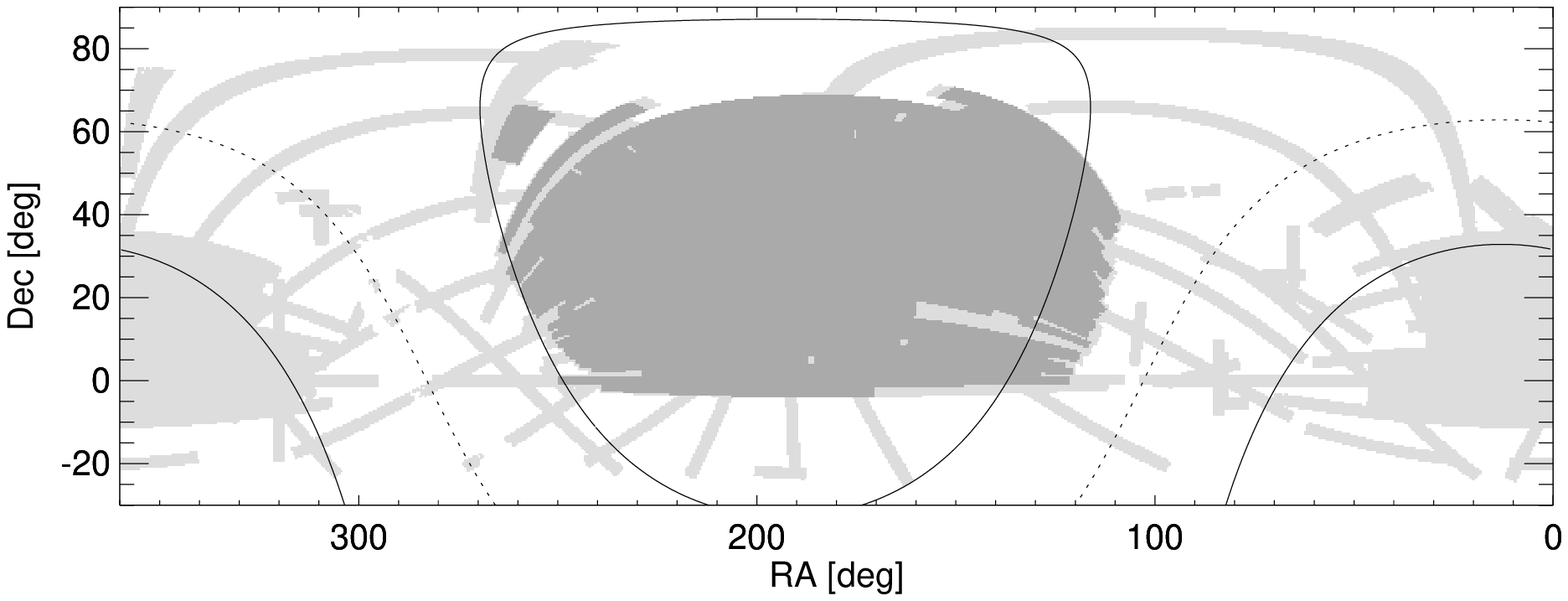}
    \end{minipage}\hfill
  \begin{minipage}{\linewidth}
    \centering
    \includegraphics[width=15cm,height=6cm]{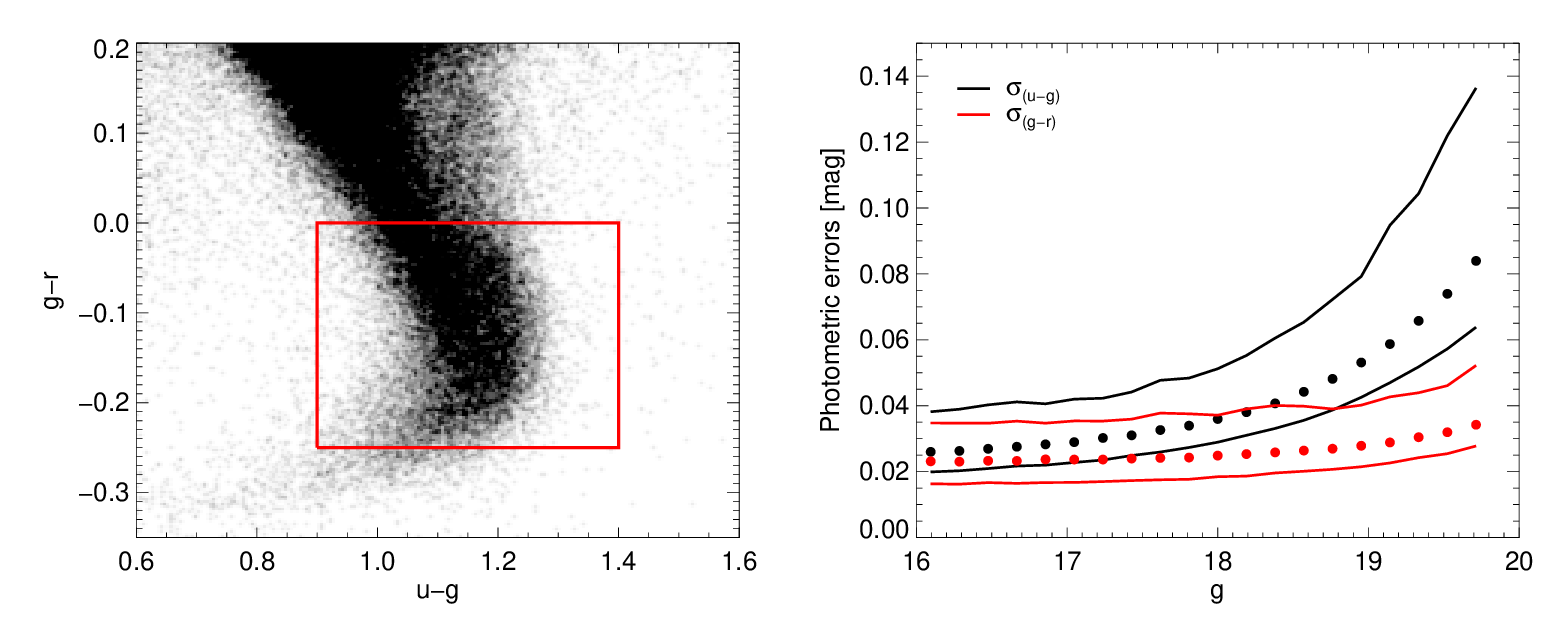}
  \end{minipage}
  \caption{\small Top panel: The SDSS DR8 footprint in equatorial
    coordinates. The solid and dotted lines show $|b| =30^\circ$ and
    $b=0^\circ$ respectively. The darker grey area shows the sky
    coverage of the SDSS data release 5 (DR5). The more recent DR8
    sample has both Northern and Southern sky coverage. Bottom left panel: The colour
    selection in $\umg$ and $\gmr$ used to select BHB and BS
    stars. Our sample consists of $N=20290$ stars in the magnitude
    range $16<g<18.5$. Bottom right panel: The median (dots), 5th and
    95th percentiles of the photometric error in $\umg$ (black) and
    $\gmr$ (red) colours as a function of $g$ band magnitude. Beyond
    $g\sim 18.5$, the errors in $\umg$ increase fairly rapidly.}
 \label{fig:dr8}
\end{figure*}

In the top panel of Fig. \ref{fig:dr8}, we show the sky coverage of SDSS data release 8 (DR8) in equatorial
coordinates. For comparison, the sky coverage of the SDSS data release
5 (DR5) is indicated by the darker grey region. The more recent SDSS
data release covers both Northern and Southern latitudes. We exclude latitudes $|b| < 30^\circ$, so as to
concentrate on regions well away from the plane of the galaxy. Over
the distance range probed by this work, the SDSS footprint ($|b| >
30^\circ$) covers approximately 20\% of the total volume of the
stellar halo. In this study we use blue horizontal branch (BHB) and
blue straggler stars (BS) to map the density profile of the stellar
halo. These A-type stars are selected by choosing stars in the
following region in colour-colour space:
\begin{eqnarray}
0.9 < \umg < 1.4 \notag\\
-0.25 < \gmr < 0.0 
\end{eqnarray}
This selection is similar to other work using A-type stars
(e.g. \citealt{yanny00}; \citealt{sirko04}) and is chosen to exclude
main sequence stars, white dwarfs and QSOs. The bottom left hand panel
of Fig. \ref{fig:dr8} highlights the colour selection box.  Whilst we
assume that all of our selected stars are BHBs or BSs, there may be a
non-negligible contribution by variable stars, such as RR Lyrae. We
use multi-epoch stripe 82 data to estimate the fraction of variable
stars in the same magnitude and colour range as our sample. We use the
light curve catalogue compiled by \cite{bramich08} and classify
variable stars according to the criteria outlined in
\cite{sesar07}. The resulting fraction of variable stars is $\sim
5\%$. This small fraction of non-BHB and non-BS stars will therefore
make little difference to the results of this work.

In the bottom right hand panel of Fig. \ref{fig:dr8}, we show the
error in $\umg$ and $\gmr$ colours as a function of $g$ band
magnitude. The photometric errors in $\umg$ are larger than in $\gmr$,
especially at fainter magnitudes. This can be compared with the
typical separation between BHBs and BS stars in $\umg$ colour (see
bottom panel of Figure \ref{fig:ridge}) which ranges from 0.05 to 0.1
mag. Mean photometric error $\sigma_{(\umg)}$ reaches 0.05 at $g \sim
18.5$ and beyond that the value rapidly increases. Accordingly, in
this study we only select stars in the magnitude range $16< g
<18.5$. This corresponds to a distance range of $\sim 4-40$ kpc for
typical absolute magnitudes of BHB and BS stars. Note that BHB and BS
stars have different absolute magnitudes, and so span separate, but
overlapping, distance ranges (see Section \ref{sec:absmag}).

\subsection{Ridgelines in Colour-Colour Space}
\begin{figure}
  \includegraphics[width=8cm,height=12.8cm]{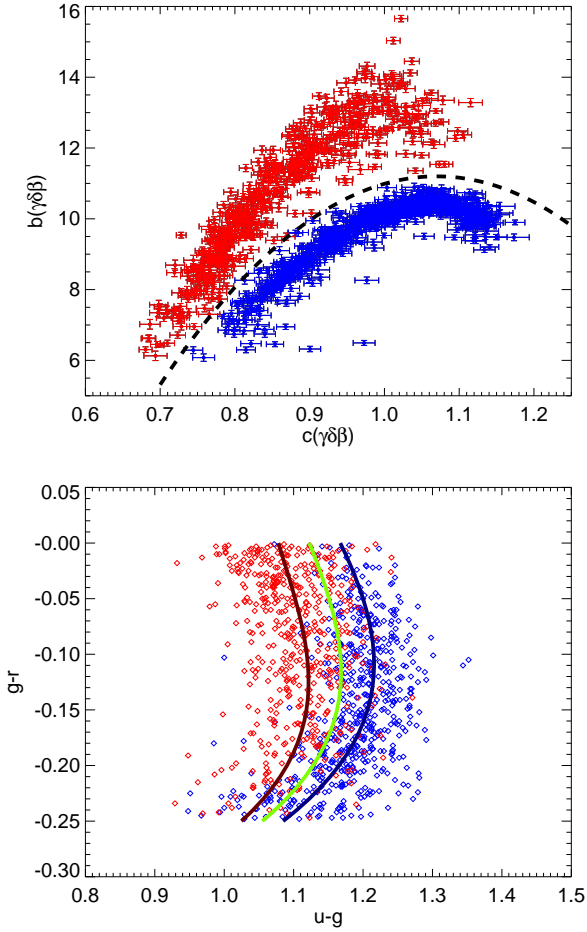}
  \caption{\small Top panel: The scale width, $b(\gamma\delta\beta)$,
    and line shape, $c(\gamma\delta\beta)$, of the Balmer lines
    $H_\gamma$, $H_{\delta}$ and $H_{\beta}$ for bright A-type stars
    taken from the SDSS data release 7 (DR7) spectral catalogue. These
    stars are selected in the same colour range as our DR8 photometric
    sample and have magnitudes in the range $16<g<17$. The blue and
    red points denote BHB and BS stars respectively. The black dashed
    line shows the apparent separation of these stars based on the
    Sersic line profiles of the Balmer series. Bottom panel: The
    `ridgelines' for the BHB and BS stars in the colour space
    ($\umg,\gmr$). The thick blue and red lines show the third order
    polynomial fits to these loci in colour space. The green line
    indicates the approximate border between the two populations.}
  \label{fig:ridge}
\end{figure}
\begin{figure}
 \includegraphics[width=8cm,height=16cm]{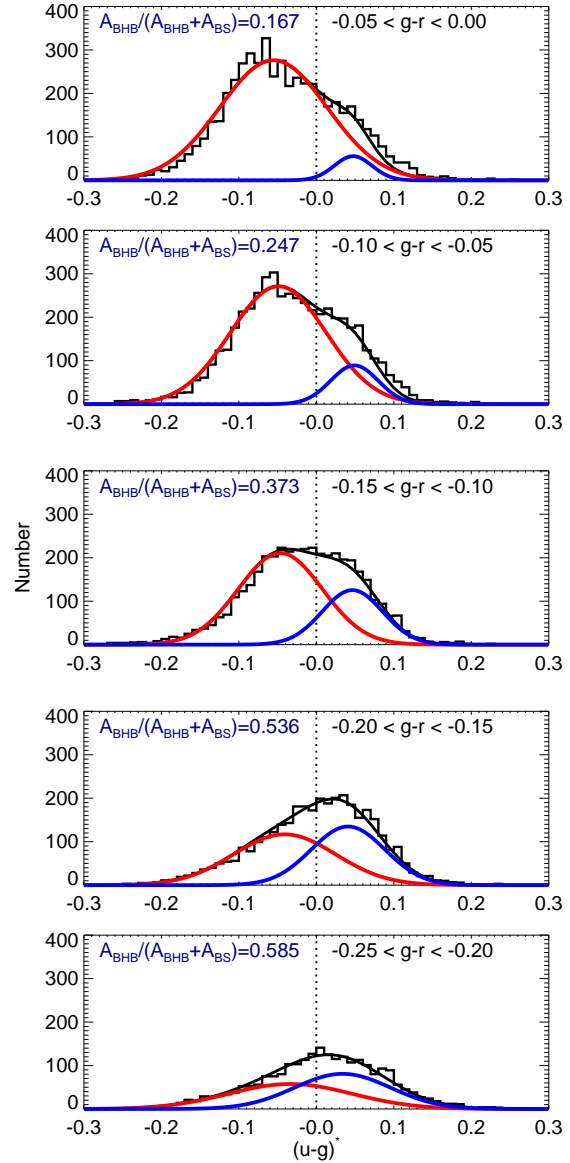}
 \caption{\small The distribution of colours in $\left(\umg
   \right)^\star$ space. Each row shows the distribution for a
   different range in $\gmr$ colour. A two component model is fit to
   the data. The overall model is shown by the black line. Individual
   Gaussians are shown by the red (BS) and blue (BHB) lines
   respectively. Stars with $\left(\umg \right)^\star < 0$ are
   dominated by BS stars whilst stars with $\left(\umg \right)^\star >
   0$ are dominated by BHB stars. The ratio between the amplitude of
   the Gaussians gives an estimate of the overall number ratio between
   the two populations.}
\label{fig:ug_star}
\end{figure}
We seek to measure the centroids of the BHB and BS loci in colour
space as well as their intrinsic widths. Naturally, this can only be
done provided there exists a robust classification of the A-type stars
according to their surface gravity. It has been shown
(e.g. \citealt{clewley02}; \citealt{sirko04}; \citealt{xue08}) that
BHB and BS stars can be separated cleanly on the basis of their Balmer
line profiles. We proceed by selecting A-type stars from the spectral
SDSS data release 7 (DR7) catalogue within the same colour range as
our DR8 photometric sample. Restricting the sample to high
signal-to-noise (S/N) spectra in the magnitude range $16 < g < 17$, we
fit the Balmer lines $H_\gamma$, $H_{\delta}$ and $H_{\beta}$ with
Sersic profile of the form,
\begin{equation}
y=1.0-a\,\exp -\left(\frac{|x-x_0|}{b}\right)^c,
\end{equation}
where $x_0$ and $a$ give the wavelength and the line depth at the line
centre respectively. The parameters $b$ (the scale width) and $c$ are
related to the line width and line shape respectively. The relation
between combined line widths and line shapes of the three Balmer lines
$H_\gamma$, $H_\delta$ and $H_{\beta}$ are shown in the top panel of
Fig. \ref{fig:ridge}. The BHBs (blue points) and BSs (red points) are
clearly separated in this diagram and the decision boundary is
indicated by the dashed black line. We use this spectral
classification to pinpoint the loci of the two populations in the
$\umg,\gmr$ colour-colour space. In our analysis, these ``ridgelines'',
i.e. approximate centres in $\umg$ as a function of $\gmr$, are defined
by third order polynomials:
\begin{eqnarray}
\label{eq:ridge}
(\umg)^{0}_{\rm BHB}&=& 1.167-0.775(\gmr)-1.934(\gmr)^2 \notag \\
&&+9.936(\gmr)^3, \notag \\
(\umg)^0_{\rm BS}&=& 1.078-0.489(\gmr)+0.556(\gmr)^2 \notag \\
&&+13.444(\gmr)^3, 
\end{eqnarray}
for $-0.25 < \gmr < 0.0$.  The ridgelines are shown by the thick blue
and red lines in the bottom panel of Fig. \ref{fig:ridge}. The green
line indicates the approximate boundary between BHBs and BSs in
$\umg$, $\gmr$ space. In addition, we calculate the intrinsic spread
of the two populations about their ridgelines. We find $\sigma_{\rm
  BHB,0}(\umg)=0.04$ and $\sigma_{\rm
  BS,0}(\umg)=0.045$. Fig. \ref{fig:ridge} makes it apparent that,
even for brightest stars, photometric information alone is not enough
to separate BHB and BS stars. Therefore, for a given star we define
the probability of BHB or BS class membership based on its distance in
colour-colour space from the appropriate locus. We assume that both
populations are distributed in a Gaussian manner about their
ridgelines. We model the conditional probability of measuring $\umg$
and $\gmr$ colours, given the star of each species, as
\begin{eqnarray}
\label{eq:prob}
p(ugr~|~{\rm
  BHB})\propto\mathrm{exp}\left(-\frac{\left[(\umg)-(\umg)_{\rm
      BHB}^0\right]^2}{2\sigma_{\rm BHB}^2}\right),
\notag\\ p(ugr~|~{\rm
  BS})\propto\mathrm{exp}\left(-\frac{\left[(\umg)-(\umg)_{\rm
      BS}^0\right]^2}{2\sigma_{\rm BS}^2}\right).
\end{eqnarray}
Note that, in fact, these probabilities are also functions of $\gmr$
since the centre of the Gaussian distribution, $(\umg)^0$ varies with
the $\gmr$ colour (see eqn. \ref{eq:ridge}). The dispersion about the
ridgeline centre depends on the intrinsic width and the photometric
errors in $\umg$, $\sigma=\sqrt{\sigma^2_{0}+\sigma_{(\umg)}^2}$. The
colour-based posterior probabilities of class membership are then
\begin{eqnarray}
\label{eq:prob2}
P({\rm BHB}~|~ugr)=\frac{p(ugr~|~{\rm BHB})~N_{\rm BHB}}{p(ugr~|~{\rm BHB})~N_{\rm BHB}+p(ugr~|~{\rm BS})~N_{\rm BS}}  \notag\\ 
P({\rm BS}~|~ugr)=\frac{p(ugr~|~{\rm BS})~N_{\rm BS}}{p(ugr~|~{\rm BHB})~N_{\rm BHB}+p(ugr~|~{\rm BS})~N_{\rm BS}}
\end{eqnarray}
The total numbers of stars $N_{\rm BHB}$ and $N_{\rm BS}$ in a given
colour range can then be found iteratively by integrating equations
(\ref{eq:prob2}). In Table~\ref{tab:fraction}, we give the fraction of
BHB and BS stars in five $\gmr$ colour bins. The fraction ranges from
$f_{\rm BHB} \sim 0.15$ at redder colours to $f_{\rm BHB} \sim 0.6$ at
bluer colours. This is in good agreement with the overall BHB to BS
ratios estimated by \cite{bell10} and \cite{xue08} who use similar
magnitude ranges.
\begin{table}
\begin{center}
\renewcommand{\tabcolsep}{0.2cm}
\renewcommand{\arraystretch}{0.5}
\begin{tabular}{| l  l  l  l |}
    \hline 
    & $N_{\rm tot}$ &  $f_{\rm BHB}$ & $f_{\rm BS}$\\
    \\
    \hline
    $-0.05 < \gmr < 0.00$ & 5189 & 0.154 &0.846  \\
    \\
    $-0.10 < \gmr < -0.05$ & 4973 & 0.231 & 0.769 \\
    \\ 
    $-0.15 < \gmr < -0.10$ & 4151 & 0.346 & 0.654 \\
    \\
    $-0.20 < \gmr < -0.15$ & 3564 & 0.536 & 0.464 \\
    \\ 
    $-0.25 < \gmr < -0.20$ & 2413 & 0.613 & 0.387   \\
    \\ 
    \hline
  \end{tabular}
  \caption{\small The fraction of BHB and BS stars in different colour
    bins. We give the $\gmr$ colour range, the total number of stars,
    the estimated fraction of BHB stars and the estimated fraction of
    BS stars.}
\label{tab:fraction}
\end{center}
\end{table}
Figure \ref{fig:ug_star} demonstrates the evolution of the separation
between the two populations in colour space. For illustration
purposes, the SDSS $\umg$ colour is transformed into
$\left(\umg\right)^\star$ using the following relation:
\begin{equation}
\left(\umg\right)^\star=(\umg)-(\umg)^0_{\rm border}
\end{equation}
Here, $(\umg)^0_{\rm
  border}=1.223-0.632(\gmr)-0.689(\gmr)^2+11.690(\gmr)^3$ is defined
by the approximate boundary line between BHB and BS stars shown by the
green line in Fig. \ref{fig:ridge}. In these new coordinates, the
curved shape of the decision boundary becomes a straight line. In
Fig. \ref{fig:ug_star}, we fit two Gaussian distributions to the
distribution in $\left(\umg \right)^\star$ for five bins in
$\gmr$. This two component model fits the overall distribution very
well. The ratio between the amplitudes of the two Gaussians varies
with $\gmr$ colour. The fraction of BHB stars increases towards bluer
colours, in good agreement with our estimates in
Table~\ref{tab:fraction}.

\subsection{Absolute Magnitudes}

\label{sec:absmag}
\begin{figure*}
  \centering
  \begin{minipage}{0.33\linewidth}
    \includegraphics[width=6.cm,height=5.cm]{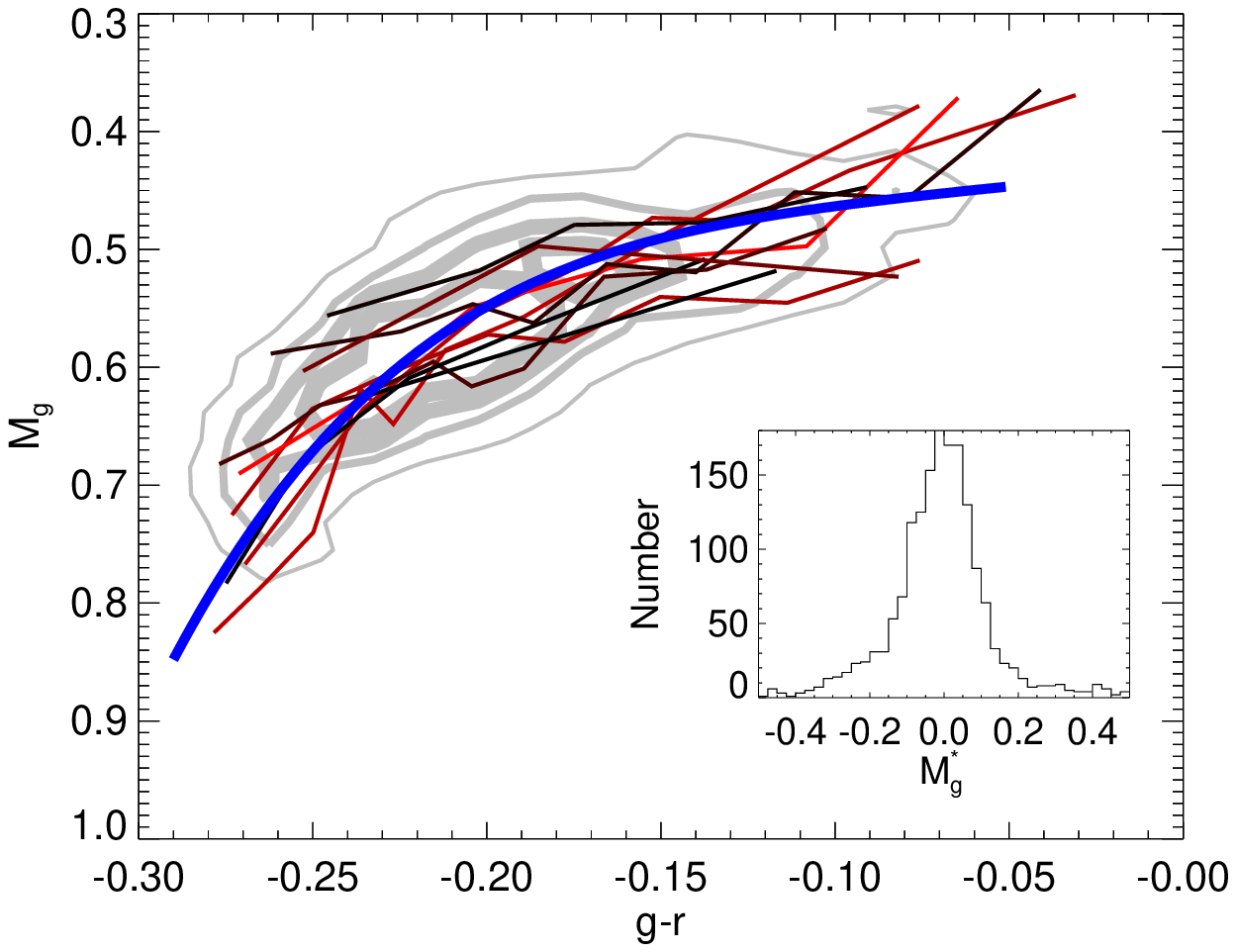}
    \end{minipage}\hfill
  \begin{minipage}{0.33\linewidth}
    \includegraphics[width=6.cm,height=5.cm]{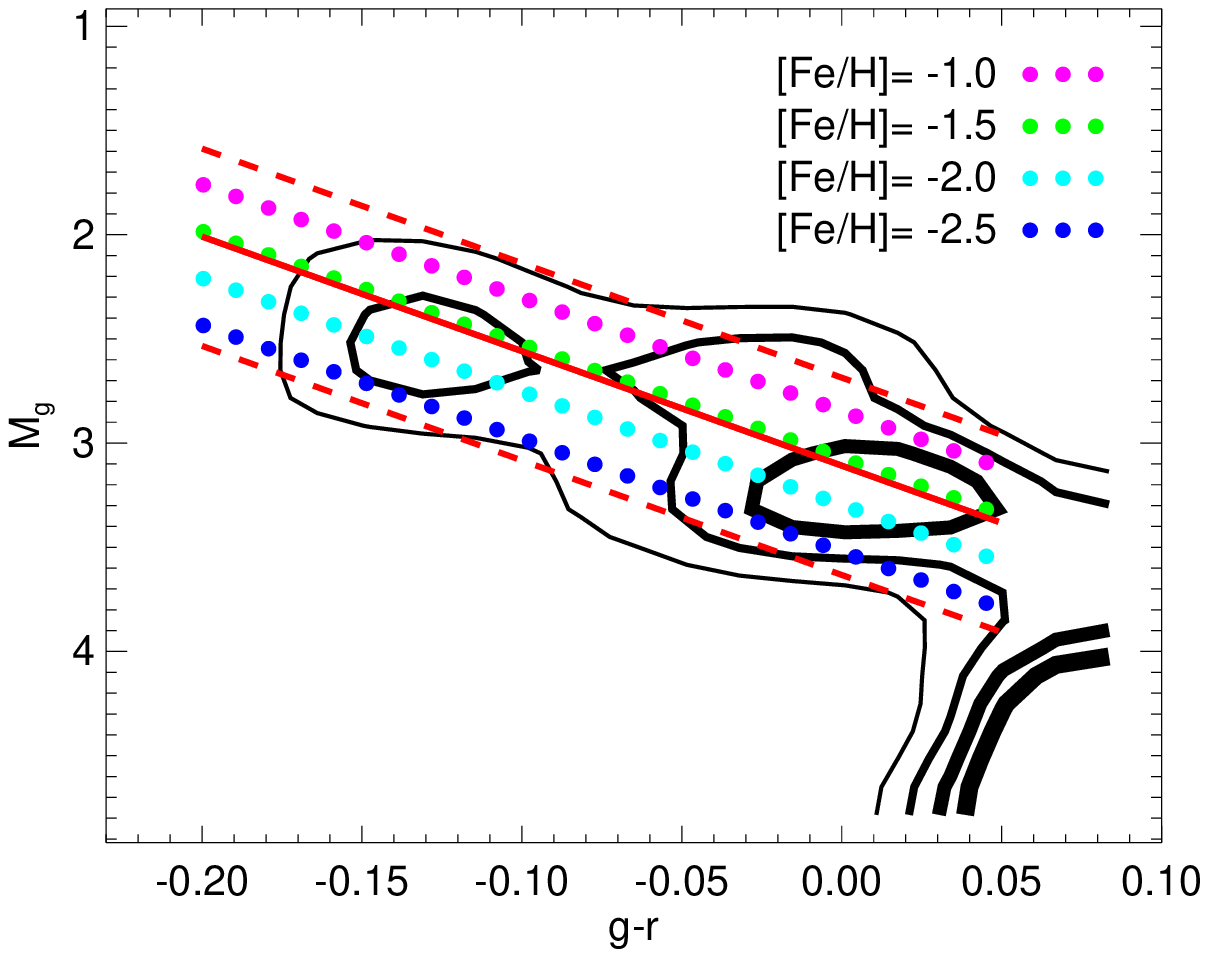}
  \end{minipage}\hfill
  \begin{minipage}{0.33\linewidth}
    \includegraphics[width=6.cm,height=5.cm]{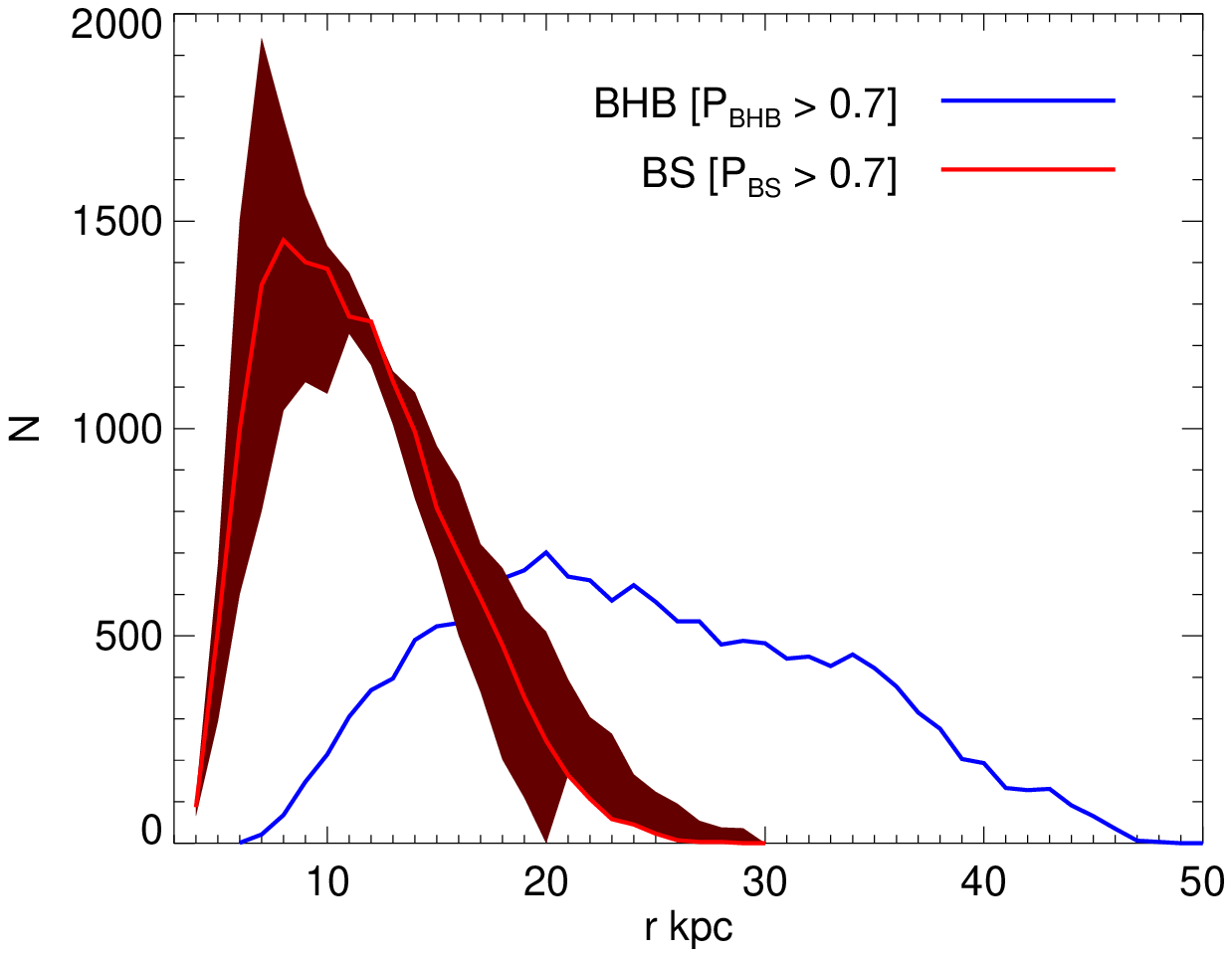}
  \end{minipage}
  \caption[]{\small Left Panel: The colour-absolute magnitude relation
    for BHB stars derived from star clusters published in
    \cite{an08}. A polynomial of order 4 is fit to the BHB stars in
    eleven star clusters: M2, M3, M5, M13, M53, M92, NGC2419, NGC4147,
    NGC5053 and NGC5466 (solid blue line). The grey contours indicate
    the density of stars within the colour-absolute magnitude region
    (thicker lines indicate higher densities). The ridgelines for
    individual clusters are shown by the red/black lines. These are
    colour coded by metallicity: red to black goes from more metal
    rich ([Fe/H] ~ -1.3) to more metal poor ([Fe/H] ~ -2.3). The inset
    panel shows the distribution of values around the derived relation
    indicating a small degree of scatter. Middle panel: The density of
    BS stars in Stripe 82 belonging to the Sagittarius stream. The
    solid and dashed red lines give the absolute magnitude colour
    relation and the estimated dispersion ($\sigma_{M_g} \sim
    0.5$). The distance to this portion of the Sagittarius stream is
    estimated from \cite{watkins09} as $D_{\rm Sgr} =26.1 \pm 5.6$
    kpc. The coloured dots show the relation derived by
    \cite{kinman94} for different metallicity BS stars in star
    clusters. Right panel: The radial distribution of the BHB and BS
    star populations. The blue line shows the distribution for high
    probability BHB stars ($P(\mathrm{BHB}) > 0.7$). The red shaded
    region shows the distribution for high probability BS stars
    ($P(\mathrm{BS})> 0.7$) where the uncertainty in the absolute
    magnitudes is taken into account.}
  \label{fig:abs_mag}
\end{figure*}  

Let us now derive a relationship between the absolute magnitude and
colour of BHB and BS stars. BHB stars are intrinsically brighter and
their absolute magnitude varies little as a function of temperature
(colour) or metallicity. In comparison, BS are intrinsically fainter
and span a much wider range in absolute magnitude.

The absolute magnitudes of BHB stars are calibrated using star
clusters with SDSS photometry published by \cite{an08}. Ten star
clusters have prominent BHB sequences; M2, M3, M5, M13, M53, M92,
NGC2419, NGC4147, NGC5053 and NGC5466\footnote{We adopt
    distance moduli for NGC2419 and NGC4147 of 19.8 and 16.32
    respectively. These differ form the values given in Table 1 of
    \cite{an08}. We find that these revised values are in better
    agreement with the colour magnitude diagrams of the
    clusters.}. The density distribution of absolute magnitudes of
BHB stars in these clusters is shown as a function of $\gmr$ colour in
the left hand panel of Fig. \ref{fig:abs_mag}. There is little
variation of the absolute magnitude with colour for BHB stars. $M_g$
changes by 0.2 (from 0.65 to 0.45) in the $-0.25 < \gmr < 0$
range. The inset panel shows the variation of absolute magnitude about
the $M_{g(\rm BHB)}$ vs. $\gmr$ trend ($M_g^*$). The spread of this
distribution is $\sim 0.1$, indicating that there is a tight relation
describing the BHB absolute magnitude. The star clusters have
metallicities typical of halo stars and ranging from $-2.3 <
[\mathrm{Fe/H}] < -1.3$, but we find no obvious trend with
metallicity.

BS stars are less common in globular clusters than BHB stars. Instead,
to calibrate the absolute magnitudes of BS stars we make use of stars
in Stripe 82 belonging to the Sagittarius stream. The distance to the
stream in the right ascension range $25^\circ < \alpha^\circ <
40^\circ$ was estimated by \cite{watkins09} using RR Lyrae stars as
$D_{\rm Sgr} =26.1 \pm 5.6$ kpc. In the middle panel of
Fig. \ref{fig:abs_mag}, we show the absolute magnitude of stars in
Stripe 82 between $25^\circ < \alpha^\circ < 40^\circ $ as a function
of colour. The density contours are constructed by using stars outside
of the range in right ascension as a background and computing the
density contrast. An obvious plume of BS stars is apparent in the
density plot. This is shown by the contour levels extending off the
main sequence. The solid red line shows the estimated absolute
magnitude colour relation for these BS stars. For comparison, we show
the absolute magnitude versus colour relation for BS stars estimated
by \cite{kinman94}. This relation is converted from Johnson-Cousins
photometry ($UBV$) to Sloan photometry ($ugr$) using the
transformation derived in \cite{jester05}. The different coloured dots
show the relation for different metallicity stars. The absolute
magnitude calibration derived for the BS stars in the Sagittarius
stream is almost identical to the \cite{kinman94} relation for BS
stars with metallicity $[\mathrm{Fe/H}] = -1.5$. This is in good
agreement with the metallicity of the stream stars found by
\cite{watkins09} ($[\mathrm{Fe/H}] = -1.43$).

We compute the spread of absolute magnitudes for each colour bin as
$\sigma_{M_g} \sim 0.5$. This dispersion takes into account the
distance errors ($D_{\rm Sgr} =26.1 \pm 5.6$ kpc) and encompasses a
range of metallicities (see dashed red lines in
Fig. \ref{fig:abs_mag}). Hence, we conclude that our calibration for
BS absolute magnitudes does not have a strong metallicity bias. Note
that the `average' BS absolute magnitude is $\sim 2.5$, approximately
$2$ magnitudes fainter than the BHB stars. This is entirely consistent
with the absolute magnitudes of halo BS stars found by \cite{yanny00}
and \cite{clewley04}.

The resulting absolute magnitudes of BHB and BS stars as a function of
$\gmr$ colour are:
\begin{eqnarray}
\label{eq:absmag}
M_{g(\rm BHB)}&=& 0.434-0.169(\gmr)+2.319(\gmr)^2 \notag\\
&&+20.449(\gmr)^3+94.517(\gmr)^4, \notag\\
M_{g(\rm BS)}&=& 3.108+5.495(\gmr),
\end{eqnarray}
which are valid over the colour range $-0.25 < \gmr < 0.0$.  For BHB
stars, this allows us to estimate accurately the distances of BHB
candidates. For BS stars, the relation is only approximate and the
scatter for each colour interval needs to be taken into account for
all distance estimates.

In the right hand panel of Fig. \ref{fig:abs_mag}, we show the
estimated radial distributions for the BHB and BS populations. We
select high probability BHB and BS stars by using the membership
probabilities defined in eqn.~(\ref{eq:prob2}). `High' probability
BHB/BS stars are defined as those for which we are $68\%$ (or
$1\sigma$) confident of BHB/BS membership. The BHB stars probe a
radial range from $r \sim 10$ kpc to $r \sim 45$ kpc. BS stars, which
have fainter absolute magnitudes, only probe out to $\sim 30$
kpc. However, there is a large degree of overlap between the two
populations between $10$ and $30$ kpc.


\section{Maximum Likelihood Method}

\label{sec:method}

In this section, we outline the maximum likelihood method used to
constrain the density profile of the stellar halo. An important
assumptions in the modelling is that the BHB and BS stars follow the
same density distribution, modulo an overall scaling. The number of
BHB stars and BS stars in a given increment of magnitude and area on
the sky is then described by
{\setlength\arraycolsep{0.1em}
\begin{eqnarray}
\label{eq:probglb}
\Delta N_{\rm BHB}(m_g\!-\!M^{\rm BHB}_g, \ell, b) &=& \rho^{0}_{\rm
  BHB}\rho(m_g\!-\!M^{\rm BHB}_g,\ell, b) D^3_{\rm BHB} \notag \\
&& \quad \times \, \frac{1}{5}\mathrm{ln}10 \, \Delta m_g \, \mathrm{cos}b \, \Delta \ell \, \Delta b \notag \\
\Delta N_{\rm BS}(m_g\!-\!M^{\rm BS}_g,  \ell,b) & =&
\rho^{0}_{\rm BS}\rho(m_g\!-\!M^{\rm BS}_g,\ell, b) D^3_{\rm BS} \\
&& \quad \times \, \frac{1}{5}\mathrm{ln}10 \, \Delta m_g \, \mathrm{cos}b \, \Delta \ell \, \Delta b . \notag
\end{eqnarray}
}
Here, the distance increment $\Delta D$ has been converted into the
apparent magnitude increment via the relation $\Delta
D=\frac{1}{5}\mathrm{ln}10 \, D \Delta m_g$. The normalising factors,
$\rho^{0}_{\rm BHB}=N_{\rm BHB}/V_{\rm BHB}$ and $\rho^{0}_{\rm
  BS}=N_{\rm BS}/V_{\rm BS}$ are found by performing volume integrals
over the SDSS DR8 sky coverage and over the required magnitude range
{\setlength\arraycolsep{0.1em}
\begin{eqnarray}
V_{\rm BHB}(M^{\rm BHB}_g)&=&\int \int \int  \left[ \rho(m_g-M^{\rm
    BHB}_g, \ell,
b) D_{\rm BHB}^3 \right. \notag \\
&& \qquad \qquad \qquad \left.  \times \, \frac{1}{5}\mathrm{ln}10 \, \mathrm{d}m_g \, \mathrm{cos}b \, \mathrm{d}\ell \, \mathrm{d}b \right]\notag \\
V_{\rm BS}(M^{\rm BS}_g)&=&\int \int \int  \left[ \rho(m_g-M^{\rm BS}_g,\ell, b)
D_{\rm BS}^3 \right. \notag \\
&& \qquad \qquad \qquad \left. \times \, \frac{1}{5}\mathrm{ln}10 \, \mathrm{d}m_g \, \mathrm{cos}b\, \mathrm{d}\ell \, \mathrm{d}b \right].
\end{eqnarray}
}
These normalising integrals depend on the absolute magnitude of the
stars and hence on the $\gmr$ colour ($M_g=M_g(\gmr)$ from eqn.
\ref{eq:absmag}). Note that the values of $V_{\rm BHB}$ and $V_{\rm
  BS}$ play a minor role in identifying the maximum likelihood model
given the choice of parameters, but are important when evaluating the
performances of different model families. We also assume that our sample consists only of BHB and
BS stars so the total number of stars is the sum of these two
populations, $N_{\rm tot}=N_{\rm BHB}+N_{\rm BS}$ where $N_{\rm
  BHB}=f_{\rm BHB}N_{\rm tot}$ and $N_{\rm BS}=f_{\rm BS}N_{\rm
  tot}$. The overall fraction of BHB to BS stars varies as a function
of $\gmr$ colour, as shown in the previous section.

Combining equations (\ref{eq:prob}) and (\ref{eq:probglb}) gives the
number of stars in a cell of colour, magnitude, longitude and latitude
space
{\setlength\arraycolsep{0.1em}
\begin{eqnarray}
 \Delta N &=& p(ugr~|~{\rm BHB})\Delta N_{\rm BHB}+p(ugr~|~{\rm
   BS})\Delta N_{\rm BS} \notag \\
&=&N_{\rm tot} \nu (ugr,l,b) \Delta (\gmr)
 \Delta m_g \Delta \ell \Delta b \mathrm{cos}b \frac{1}{5}
 \mathrm{ln}10,
\end{eqnarray}
}
where the stellar density is
{\setlength\arraycolsep{0.1em}
\begin{eqnarray}
  \nu (ugr,l,b) &=& p(ugr~|~{\rm BHB})\frac{f_{\rm BHB}}{V_{\rm BHB}}\rho(m_g\!-\!M^{\rm BHB}_g,\ell, b) D^3_{\rm BHB} +
  \notag\\ 
&& p(ugr~|~{\rm BS})\frac{f_{\rm BS}}{V_{\rm  BS}}\rho(m_g\!-\!M^{\rm BS}_g, \ell, b) D^3_{\rm BS}
\end{eqnarray}
}
Here, each star is assigned a `BHB distance' ($D_{\rm BHB}$)
\textit{as well as} a `BS distance' ($D_{\rm BS}$). The colour
probability functions weight the contribution of each star to the BHB
density or the BS density. For simplicity, we group all the stars into
five bins in $\gmr$ of width 0.05 mags. Thus, stars in each $\gmr$ bin
have the same normalisations ($V_{\rm BHB}$, $V_{\rm BS}$), fraction
of BHB/BS stars ($f_{\rm BHB}$, $f_{\rm BS}$) and absolute magnitudes
($M^{\rm BHB}_g$, $M^{\rm BS}_g$). We take into account the
uncertainty in the BS absolute magnitude by convolving the BS number
density with a Gaussian magnitude distribution. This distribution is
centred on the estimated absolute magnitude ($M^{\rm BS}_g=M^{\rm
  BS}_g(\gmr)$) and has a dispersion of $\sigma_{M_g}=0.5$ (see
Fig. \ref{fig:abs_mag}).

The log-likelihood function can then be constructed from the density
distribution,
\begin{equation}
  \mathrm{log}\mathcal{L}=\sum_{i=1}^{N_{\rm tot}} \mathrm{log} \, \left [ \nu ( m^i_g,ugr^i,\ell^i,b^i)~\mathrm{cos}b \right ].
\end{equation}
The number of free parameters constrained by the likelihood function
depends on the complexity of the model stellar-halo density
profile. The log-likelihood is maximised to find the best-fit
parameters using a brute-force grid search.

%
\begin{table}
\begin{center}
\renewcommand{\tabcolsep}{0.12cm}
\renewcommand{\arraystretch}{1.3}
\begin{tabular}{| l  c  c  l  l   r |}
    \hline 
    $N_{\rm tot}$ & $\alpha$ &  $q$ &$\mathrm{ln}(\mathcal{L}) \times 10^{4}$ & $\sigma/\rm
    tot$ & with V\&S?
    \\
    \hline
    20290 & $2.60^{+0.05}_{-0.05}$ & $0.65^{+0.02}_{-0.02}$ & -17.1171
    & $0.38 \pm 0.01$ & yes
    \\
    15403 & $2.90^{+0.05}_{-0.1}$ & $0.53^{+0.02}_{-0.01}$ & -12.2917
    & $0.22 \pm 0.02$ & no
    \\ 
    \hline
  \end{tabular}
  \caption{\small A summary of our best-fit oblate power-law models
    with and without the Virgo Overdensity and the Sagittarius
    stream. We give the total number of stars, the model parameters,
    the average log-likelihood value for the model and $\sigma/\rm
    tot$.}
\label{tab:res}
\end{center}
\end{table}
\section{Results}
\label{sec:results}
In this section, we outline the results of applying our maximum
likelihood method to the sample of A-type stars selected from the SDSS
DR8. We consider in turn a number of simple density profiles with
constant flattening -- single power-law, broken power-law and Einasto
-- before examining the case for refinements, such as triaxiality,
radial variations in shape and substructure.

\begin{figure*}
  \centering
  \begin{minipage}{0.48\linewidth}
    \centering
    \includegraphics[width=9cm,height=7cm]{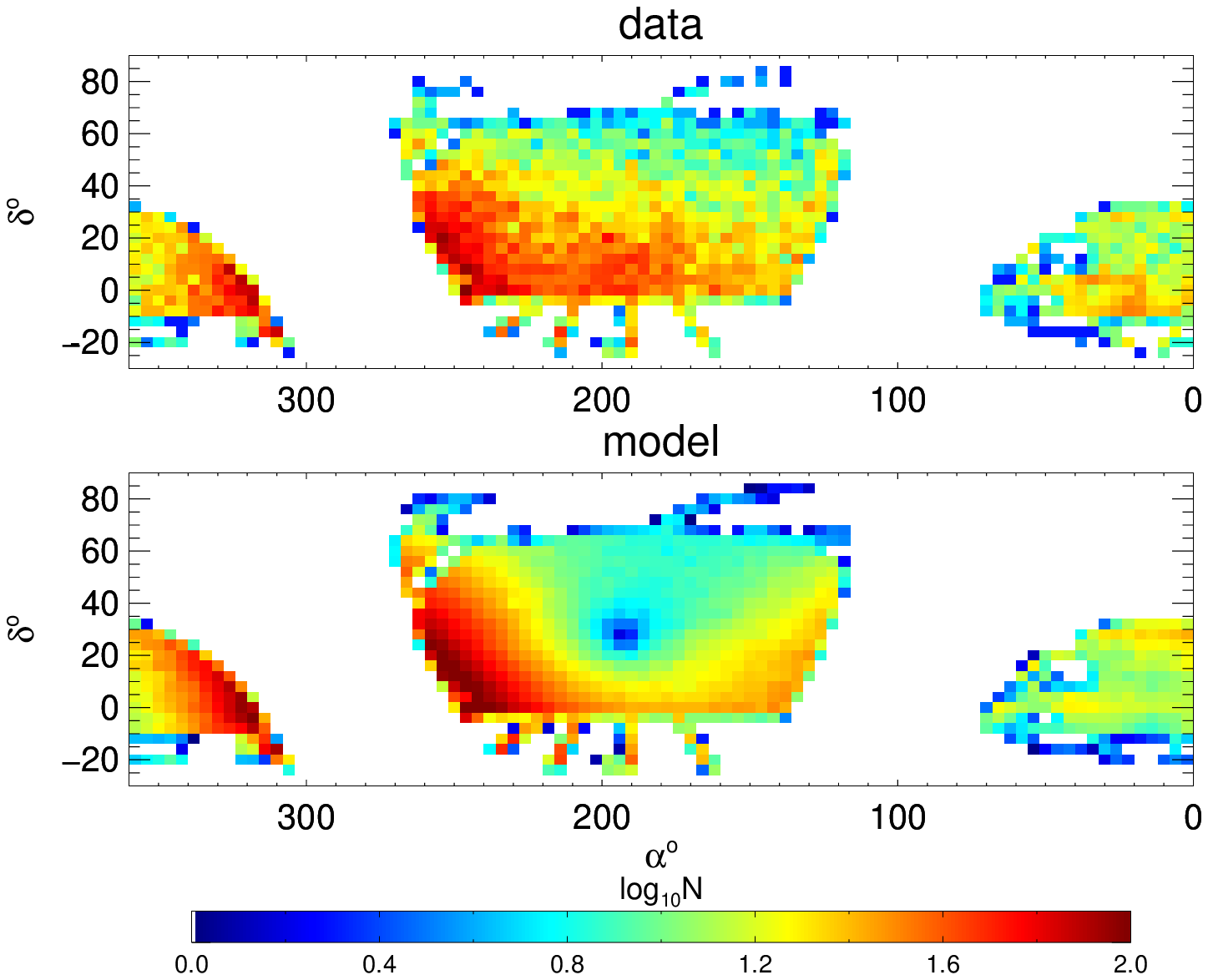}
    \end{minipage}\hfill
   \begin{minipage}{0.48\linewidth}
    \centering
    \includegraphics[width=9cm,height=7cm]{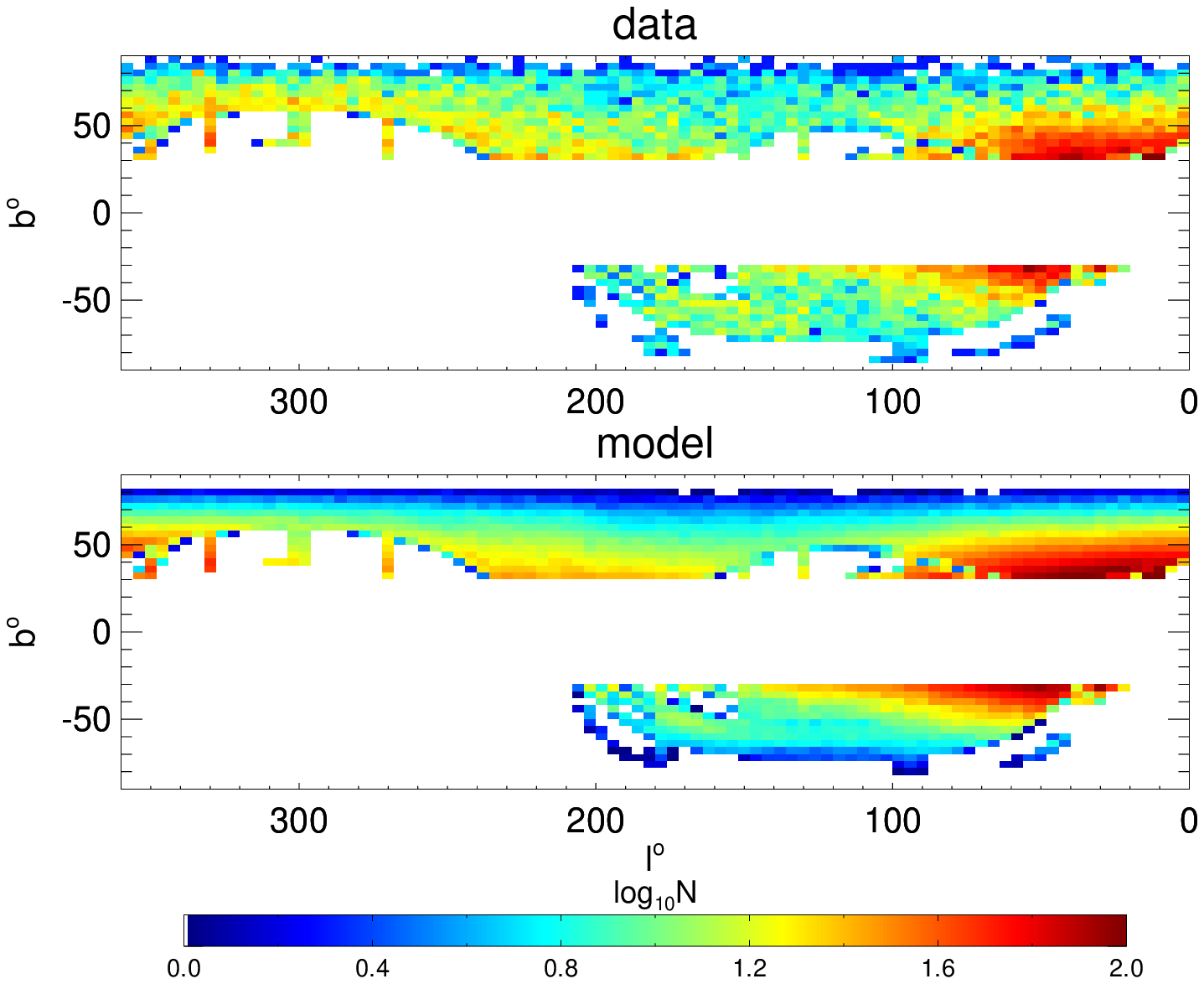}
    \end{minipage}
  \begin{minipage}{0.48\linewidth}
    \centering
    \includegraphics[width=9cm,height=4cm]{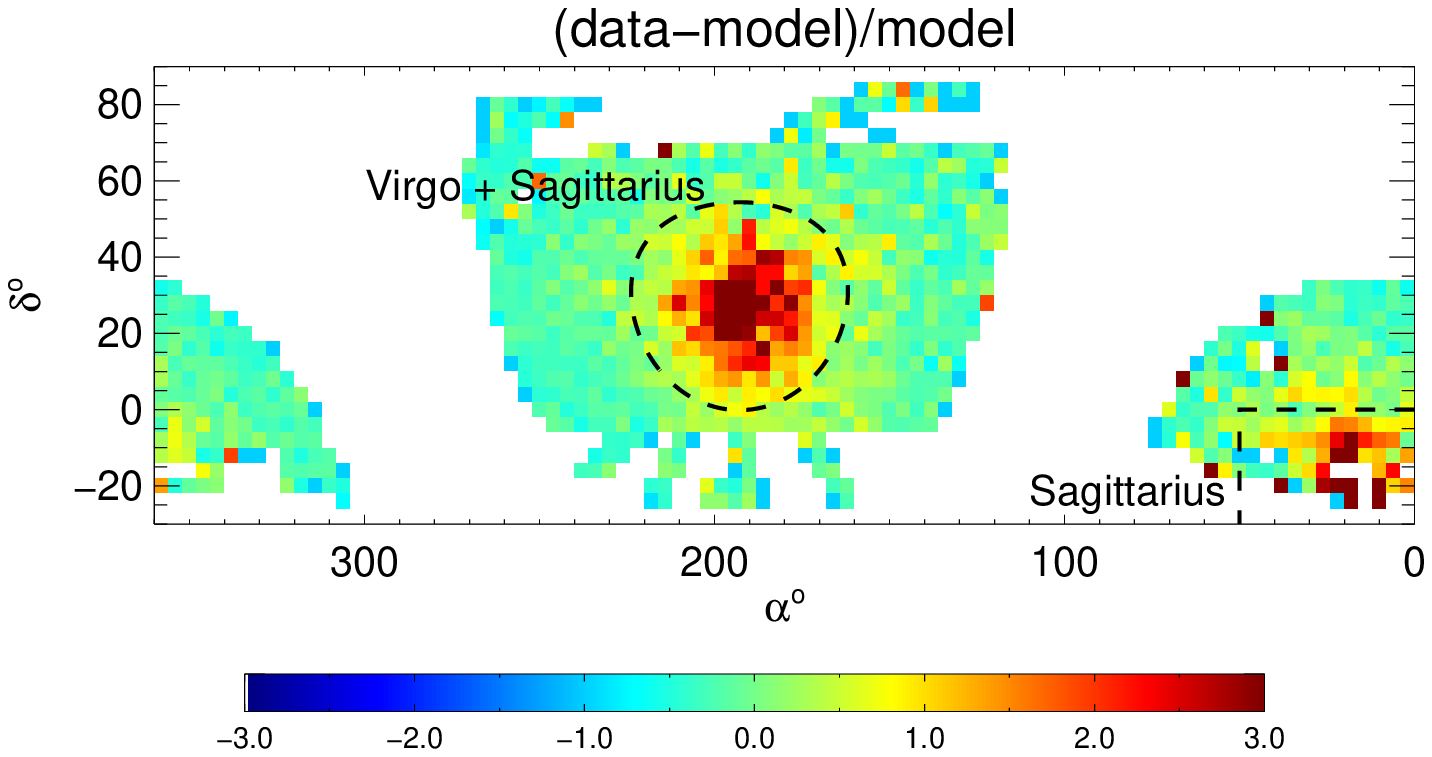}
  \end{minipage}\hfill
  \begin{minipage}{0.48\linewidth}
    \centering
    \includegraphics[width=9cm,height=4cm]{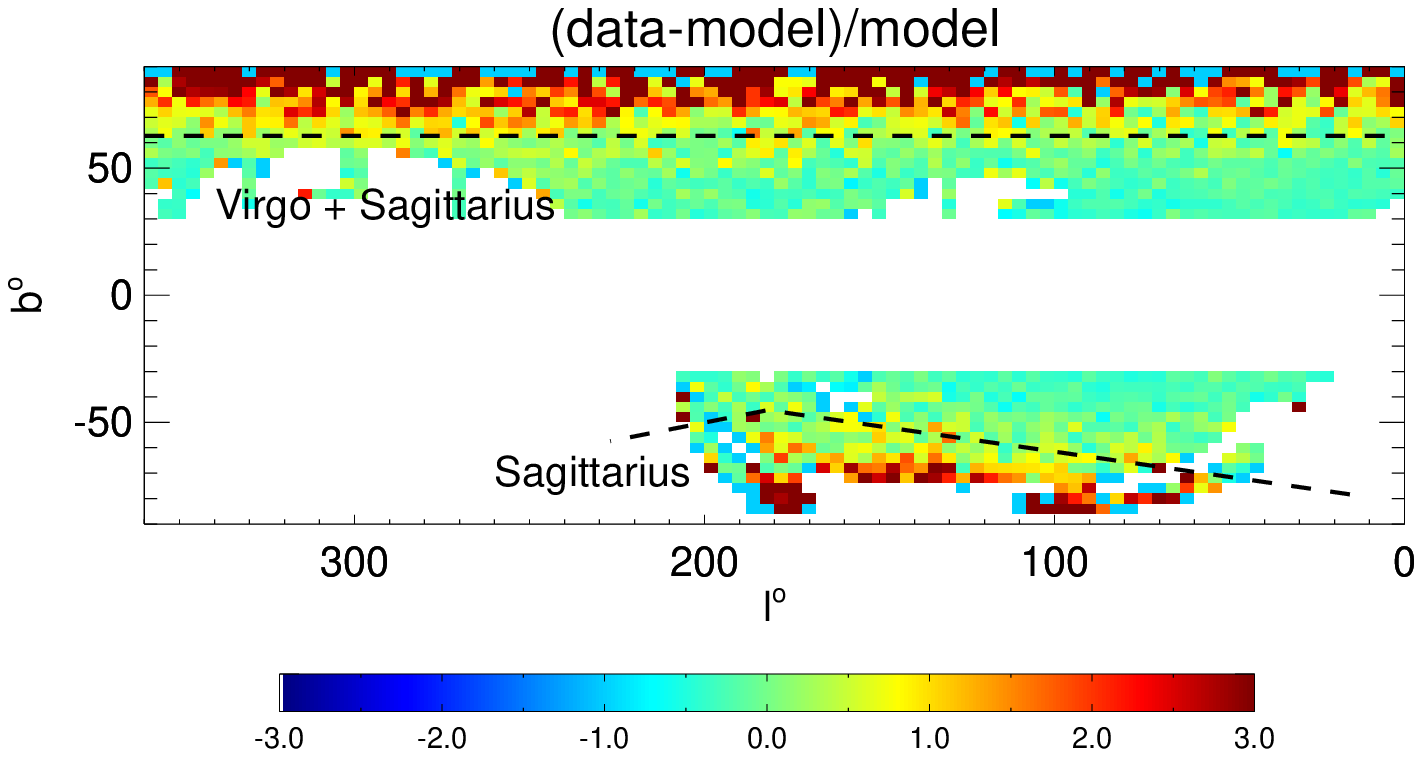}
  \end{minipage}

  \caption{\small Left panels: Density plots in equatorial
    coordinates. Right panels: Density plots in Galactic
    coordinates. The top and middle panels show the density plots for
    the the data and single power-law model respectively. The bottom panels show the
    residuals of the best-fit single power-law model. The dark red regions show the
    obvious overdensities of Virgo and Sagittarius. The dashed lines
    indicate the regions of sky removed in the maximum likelihood
    procedure. The Virgo Overdensity and parts of the leading tail of
    the Sagittarius stream are apparent in the North Galactic Cap
    whereas another portion of the Sagittarius stream is found in the
    South Galactic Cap. Note that these have been excised when
    calculating the best-fit single power-law model. The regions away from these known
    overdensities are reasonably well fit by a smooth, power-law
    density model.}
  \label{fig:residuals}
\end{figure*}  
\begin{figure}
\centering
    \includegraphics[width=8cm,height=7cm]{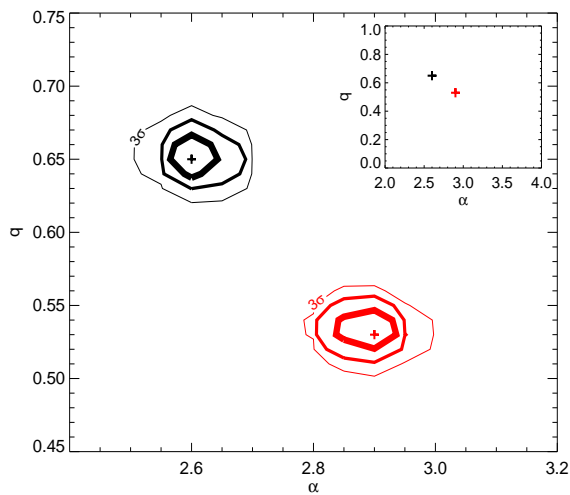}
    \caption{\small The maximum likelihood contours for the flattening
      $q$ and power-law index $\alpha$ of our single power-law halo models. The black
      (red) lines show the $1\sigma$, $2\sigma$ and $3\sigma$ contours
      when stars in the region of the Virgo overdensity and
      Sagittarius stream have been included (excluded). The inset
      panel illustrates that the difference in maximum likelihood
      parameters is relatively small whether the overdensities are
      included or excluded.}
\label{fig:likelihood}
\end{figure}
\subsection{Single Power-Law Profile}

First, let us consider a simple power-law density model of the form,
\begin{equation}
  \rho(\rellip) \propto \rellip^{-\alpha},\qquad\qquad
  \rellip^2 =x^2+y^2+z^2q^{-2}
\label{eq:rn}
\end{equation}
The parameters $q$ and $\alpha$ describe the halo flattening and the
power-law fall-off in stellar density respectively.  Oblate density
distributions have $q <1$, spherical $q = 1$ and prolate $q>1$.

\begin{figure*}
\centering
\includegraphics[width=14cm,height=5cm]{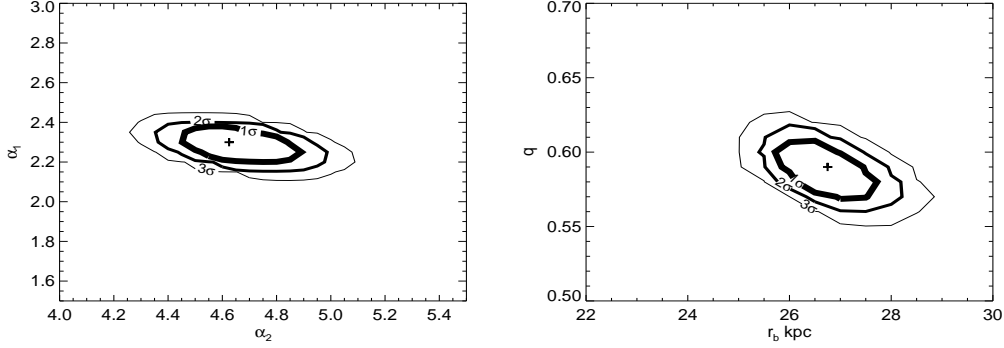}
\caption{\small Likelihood contours for the broken power-law
     models. Left panel: The contours for the inner ($\alphain$) and
     outer ($\alphao$) power laws. The fall off steepens at larger
     radii. Right panel: The contours for the break radius ($\rb$) and
flattening ($q$). The maximum likelihood solution favours a flattening
of $q=0.59$ and a break radius of $\rb=27$ kpc with inner and outer power-laws of $\alphain=2.3$ and $\alphao=4.6$ respectively. }
\label{fig:dpl}
\end{figure*}

We summarise the best-fitting single power-law model parameters in Table \ref{tab:res}.
Using the entire sample, we find the maximum likelihood model
parameters of $\alpha=2.6$ and $q=0.65$. We repeat the analysis by
excising stars in the regions of the Virgo Overdensity and Sagittarius
stream, which amounts to removing stars in the regions defined by
\begin{equation}
0 < X < 30, \quad\quad  X=63.63961\sqrt{2(1 - \sin b)}.
\end{equation}
This is the mask introduced by \cite{bell08} to remove stars belonging
to the Virgo overdensity (as well as parts of the leading tail of the
Sagittarius stream). Another portion of the Sagittarius stream is
located in the Southern part of the sky (see Fig. \ref{fig:residuals})
and is removed by
\begin{equation}
  0^\circ < \alpha < 50^\circ, \quad
  -30^\circ < \delta < 0^\circ .
\end{equation}
By discarding these large overdensities, we find a slightly steeper
power law and a more flattened shape with $\alpha=2.9$ and
$q=0.53$. There is also a substantial increase in the maximum
likelihood, implying that the fit is indeed affected by the presence
of these large overdensities. In Fig ~\ref{fig:likelihood}, we show
the maximum likelihood contours for the model parameters $q$ and
$\alpha$ for both cases. The contours encompass the $1\sigma$,
$2\sigma$ and $3\sigma$ regions respectively. We find that our
likelihood function has a well defined peak. The maximum likelihood
model parameters ($\alpha=2.9$, $q=0.53$ excluding overdensities) are
in good agreement with some of the previous work assuming a single
power-law model for the stellar halo (e.g. \citealt{yanny00};
\citealt{newberg06}; \citealt{juric08}). This exercise indicates that
the two large overdensities induce a bias in deduced model parameters
and, therefore, in the subsequent sections, we remove stars in the
vicinity of the Virgo Overdensity and Sagittarius stream in all our
calculations.

We use our maximum likelihood smooth oblate halo model ($\alpha=2.9$,
$q=0.53$) to show the residuals of the data minus model on the sky. In
Fig. \ref{fig:residuals}, we compare the data and model in both
equatorial and galactic coordinates. The top and middle panels show
the data and model on the sky (the magnitude distribution has been
collapsed) whilst the bottom panels show the residuals of the
model. The Virgo overdensity is the most obvious feature located at
($\alpha \sim 190^\circ$, $\delta \sim 30^\circ$). This overdensity
covers a substantial fraction of the North Galactic cap. A portion of
the Sagittarius stream can be seen at ($\alpha \sim 20^\circ$, $\delta
\sim -25^\circ$) in the South Galactic cap. The residuals for these
features reach up to approximately three times the model values. We
can estimate the total fraction of stars residing in these
overdensities from the excess numbers of stars in these regions of the
sky. Approximately $\sim 5.5\%$ of our total sample ($N_{\rm
  tot}=20290$) reside in the Northern Virgo+Sagittarius overdensity
whilst less than $1\%$ occupy the Southern portion of the Sagittarius
stream. The fraction of stars in these overdensities is relatively
small as they are located in the vicinity of the Northern and Southern
Galactic caps where the density of stars is small. However, as the
relative difference (i.e. (data-model)/model) between the data and
model is large in these regions, these overdensities can influence the
likelihood values.

\subsection{Broken Power-Law Profile}

We relax our models to allow a change in the steepness of the density
fall off and consider broken power-law models of the form,
\begin{equation}
\rho(\rellip) \propto \begin{cases} \rellip^{-\alphain} & \rellip \le \rb  \\
                              \rellip^{-\alphao} & \rellip > \rb. 
\end{cases}
\end{equation}

In Fig. \ref{fig:dpl} we show the maximum likelihood contours for the
inner and outer power-laws (left hand panel) and break radii and
flattening (right hand panel). A model with break radius $\rb=27$ kpc
is preferred with slopes of $\alphain=2.3$, $\alphao=4.6$
respectively. A steeper power-law is favoured at larger radii while
the power-law within the break radius is shallower. Note that the
break radius is an ellipsoidal distance and only corresponds to a
radial Galactocentric distance in the Galactic plane ($z=0$). The
model is slightly less flattened ($q \sim 0.6$) than the single
power-law model but, even with an additional two parameters, an oblate
halo model is favoured.

The maximum likelihood value for a broken power-law model is
significantly larger than for a single power-law model
($-2\mathrm{ln}(\mathcal{L}_{\rm SPL}/\mathcal{L}_{\rm BPL}) \sim
400$) (see Table \ref{tab:like}). Our broken power-law model is in
very good agreement agreement with the results of \cite{watkins09},
who use a sample of RR Lyrae stars to probe the density distribution
of the stellar halo out to $\sim 100$ kpc (see also
\citealt{sesar10}).

\begin{figure*}
\centering
\includegraphics[width=16cm,height=10cm]{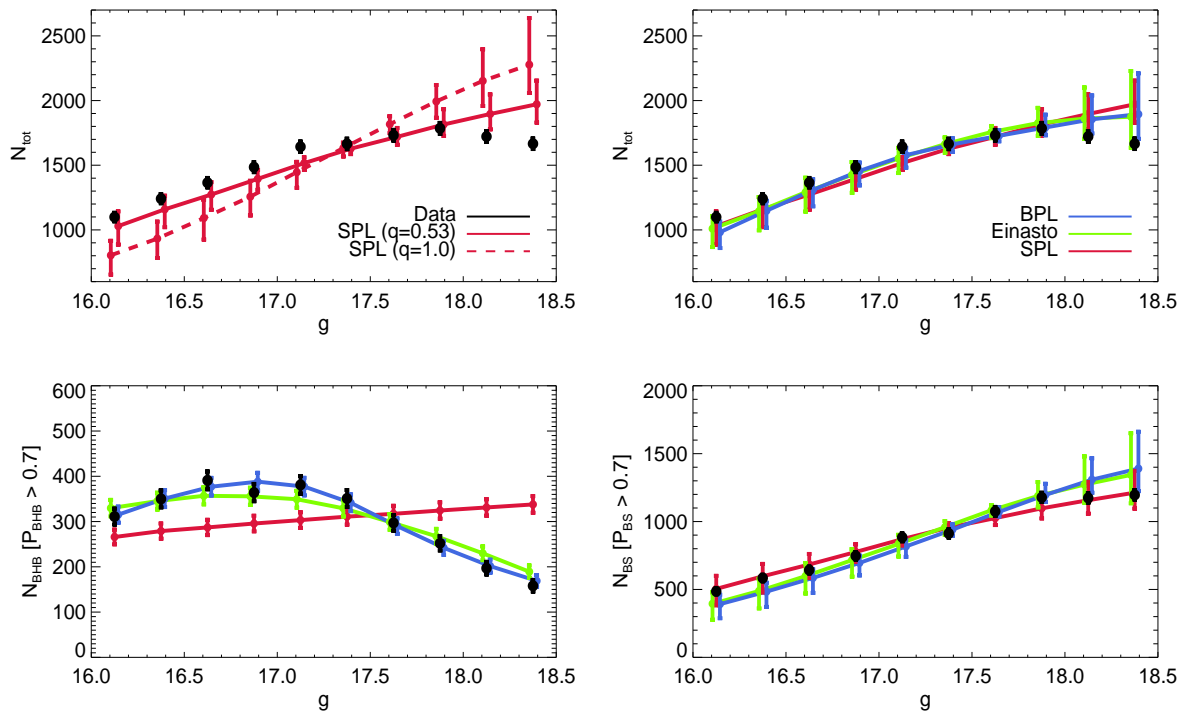}
\caption{\small The magnitude distribution of our DR8 data sample. The
  black points give the distribution of the data where the error bars
  are Poissonian. Top-left panel: Single power-law models. The solid
  and dashed red lines show the best fit oblate and spherical models
  respectively. Error bars incorporate Poissonian uncertainties and
  the spread of absolute magnitudes for BS stars. A flattened model
  provides a much better fit to the data. Top-right panel: The solid
  red, blue and green lines show the distributions for the best-fit
  single power-law, broken power-law and Einasto models
  respectively. The latter two provide a better representation of the
  data. Bottom-left panel: The magnitude distribution for the most
  probable BHB stars (with $P_{\rm BHB} > 0.7$). Bottom-right panel:
  The magnitude distribution for the most probable BS stars (with
  $P_{\rm BS} > 0.7$). }
\label{fig:gmag}
\end{figure*}

\subsection{Einasto Profile}

The Einasto profile (\citealt{einasto89}) is often used to describe
the density distribution of dark matter haloes
(e.g. \citealt{graham06}; \citealt{merritt06};
\citealt{navarro10}). The Einasto model is given by the equation
\begin{equation}
\mathrm{ln} \left[\rho(\rellip)/\rho(\reff)\right] = -d_n[(\rellip/\reff)^{1/n}-1].
\end{equation}
The shape of the density profile is described by the parameter
$n$. Density distributions with steeper inner profiles and shallower
outer profiles are generated by large values of $n$. For example, dark
matter haloes with `cuspy' inner profiles typically have values of $n
\sim 6$.  The parameter $d_n$ is a function of $n$. For $n \ge 0.5$ a
good approximation is given by $d_n= 3n-1/3+0.0079/n$
(\citealt{graham06}).

This profile allows for a non-constant fall-off without the need for
imposing a discontinuous break radii. We find the maximum likelihood
solutions for $q$, $n$ and
$\reff$. Our best-fit Einasto model has parameters $q=0.58$, $n=1.7$ and $\reff =20$
kpc. The slope of the density profile varies rapidly with radius as
indicated by the relatively small value of $n$.

\subsection{Model Comparisons}

We now test how accurately our models represent the observed magnitude
distribution of the data.  Using Monte Carlo methods, we create a
distance distribution according to the model density profile. The fake
data is given a $\gmr$ colour distribution drawn from the real data,
which can then be converted into absolute magnitudes (see
eqn. \ref{eq:absmag}). The ratio of BHB and BS stars in each colour
bin is chosen to match the values given in Table
\ref{tab:fraction}. In the case of the BS stars, absolute magnitudes
are determined from the $\gmr$ colour by drawing randomly from a
Gaussian distribution centered on the estimated value (from
eqn. \ref{eq:absmag}) with a dispersion of 0.5 mags. The resulting
(apparent) magnitude distributions for our models are compared with
the data in Fig. \ref{fig:gmag}

In the top-left hand panel, we show the magnitude distribution for our
single power-law oblate model with the solid red line. The error bars
take into account the Poisson uncertainty as well as the uncertainty
spread of absolute magnitudes for BS stars. The distribution of
magnitudes for our DR8 A-type star sample is shown by the black
points. There is reasonably good agreement but there is a notable
deviation at fainter magnitudes. For comparison, we show the maximum
likelihood \textit{spherical} model by the red dashed line (with
$\alpha \sim 2.7$), which provides a very poor fit to the data. The
top-right hand panel shows the magnitude distributions for the single
power-law, broken power-law and Einasto profiles by the solid red,
blue and green lines respectively. The broken power-law and Einasto
models provide a better representation of the data than a single
power-law, although there are still discrepancies at the faintest
magnitudes.

In the bottom panels of Fig. \ref{fig:gmag}, we show the magnitude
distributions for the most probable BHB and BS stars. We select stars
with $P_{\rm BHB} > 0.7$ as `BHB' stars and stars with $P_{\rm BS} >
0.7$ as `BS' stars (see section \ref{sec:absmag}). Of course, this is
for illustration purposes only, as a clean separation of BHB and BS
stars requires spectroscopic classification. While the BS stars are
adequately described by a single power-law model, this is a poor
description of the BHB stars. This is unsurprising, as the BS stars
cover a smaller distance range and barely populate distances beyond
the break radius. The bottom-left panel clearly shows the need for a
more steeply declining density law at larger radii.

\begin{figure*}
\centering
\includegraphics[width=14cm,height=5cm]{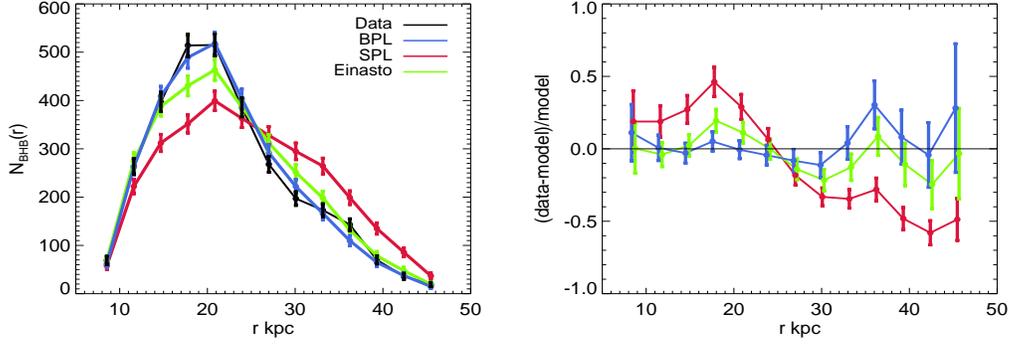}
\caption{\small Left panel: The number of BHB stars in spherical
  shells.  We select stars with $P(\rm BHB) > 0.7$ as `BHB' stars and
  show their radial distribution with the thick black line. The solid
  blue, red and green lines show our best-fit broken power-law, single
  power-law and Einasto models respectively. Right panel: The
  residuals for our best fit models.}
\label{fig:n_r}
\end{figure*}

In Fig. ~\ref{fig:n_r}, we show the distribution of (probable) BHB
stars in spherical shells with the solid black line. Here, we only
consider the most likely BHB stars (with $P_{\rm BHB} > 0.7$) as for
these we can accurately estimate their distance.  The red, blue and
green lines show the radial distributions for our best fit single
power-law, broken power-law and Einasto models respectively. The
single power-law is a poorer description of the data, whilst the
broken power-law and Einasto models both provide better
representations of the radial distribution.

\begin{table*}
\begin{center}
\renewcommand{\tabcolsep}{0.1cm}
\renewcommand{\arraystretch}{1.3}
\begin{tabular}{| l  l  c  c  c  c  c  c |}
    \hline 
    Model & Parameters & $N_p$ & $\mathrm{ln}(\mathcal{L}) \times 10^{4}$ &
    $-2\mathrm{ln}(\mathcal{L}_{\rm max}/\mathcal{L})$ &$\mathrm{ln}(E/E_{\mathrm{max}})$ & $\sigma/\rm tot$ (w/o V \& S) &  $\sigma/\rm tot$ (V \& S)\\
    \hline
    SPL - spherical & $\alpha=2.7^{+0.05}_{-0.05}$, $\mathbf{q=1.0}$ &
    1 &  -12.5516 & 5606 & -2801 & $0.44 \pm 0.01$ & $0.49 \pm 0.01$
    \\
    SPL - oblate & $\alpha=2.9^{+0.04}_{-0.06}$,
    $q=0.53^{+0.02}_{-0.01}$ & 2 & -12.2917 & 408 & -206 & $0.22 \pm 0.02$
    & $0.38 \pm 0.01$
    \\
    SPL - triaxial & $\alpha=2.9^{+0.05}_{-0.05}$,
    $q=0.50^{+0.02}_{-0.01}$, $p=0.71^{+0.03}_{-0.03}$ & 3 & -12.2818
    & 210 & -110 & $0.20 \pm 0.02$ & $0.34 \pm 0.01$
    \\ 
    SPL - $q=q(r)$ & $\alpha=2.9^{+0.05}_{-0.05}$,
    $q=0.53^{+0.02}_{-0.01}$, $r_0 > 10^3$kpc & 3 & -12.2917 & 368 &
    -206 &
    $0.22 \pm 0.02$ & $0.38 \pm 0.01$
    \\ 
    BPL - oblate & $r_{b}=27^{+1}_{-1}$kpc,
    $q=0.59^{+0.02}_{-0.03}$, & 4 & -12.2713 & 0  & 0 &$0.21 \pm 0.02$ & $0.36
    \pm 0.01$
    \\
    & $\alphain=2.3^{+0.1}_{-0.1}$, $\alphao=4.6^{+0.2}_{-0.1}$ & & &
    & & &
    \\
    Einasto - oblate & $n=1.7^{+0.2}_{-0.2}$,
    $r_{\rm eff}=20^{+1.0}_{-1.0}$kpc, $q=0.58^{+0.02}_{-0.02}$ & 3 & -12.2757 & 88
    & -45 & $0.22 \pm 0.01$  & $0.37 \pm 0.01$
    \\
    \hline
  \end{tabular}
  \caption{\small A summary of our best-fit models. We give the type
    of model, the best-fit parameters of the model, the number of free
    parameters, the maximum log-likelihood, the difference in
    likelihood relative to the maximum likelihood model, the log
    evidence$^{1}$ ratio relative to the maximum likelihood model and
    $\sigma/\rm tot$ both with and without the Virgo and Sagittarius
    overdensities. Parameters which are kept fixed are highlighted in
    bold.}
\label{tab:like}
\end{center}
\end{table*}

\footnotetext[1]{The Bayesian evidence is the integral of the
  likelihood values over the parameter space (assuming a uniform
  prior). $E \approx \int \mathcal{L}(\theta) \mathrm{d}\theta$, where $\theta$
  is the model parameter vector.}

\subsection{Refinements: Triaxiality and a Radially Dependent Shape}

A natural question to ask is whether further refinements might provide
a still more accurate description of the data. Up to now, we have
assumed spheroidal halo models with a constant flattening with radius.

\begin{figure}
\centering
    \includegraphics[width=8cm,height=4cm]{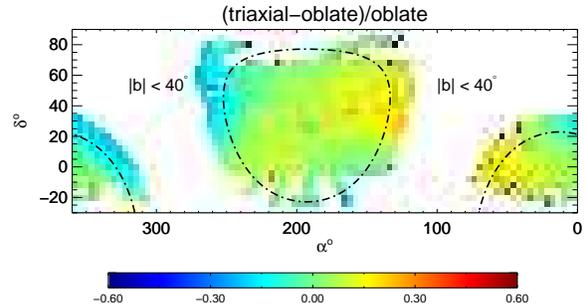}
    \caption{\small The residuals of our maximum likelihood triaxial
      and oblate models. The dot-dashed region indicates the boundary
      between low ($|b| < 40^\circ$) and high ($|b| < 40^\circ$)
      Galactic latitudes. Note the difference in scale that is used
      for this figure to that used for Fig. \ref{fig:residuals}.}
\label{fig:triax}
\end{figure}

First, we consider whether triaxiality makes any improvement. The
definition of $\rellip$ is modified to
\begin{equation}
\rellip^2 =x^2+y^2p^{-2} + z^2q^{-2}.
\end{equation}
We fit single power-law triaxial models to the data and obtain maximum
likelihood parameters of $q=0.50$ and $p=0.71$. The likelihood value
increases relative to an oblate model with the inclusion of this extra
parameter ($-2\mathrm{ln}(\mathcal{L}_{\rm oblate}/\mathcal{L}_{\rm triaxial}) \sim
200$). The magnitude distribution of the triaxial model is largely the
same as the oblate model. We inspect the difference between these two
models in equatorial coordinates in Fig. \ref{fig:triax}. The
dot-dashed lines indicates the regions of sky with low galactic
latitudes $|b| < 40^\circ$. Regions with latitudes below $|b| <
30^\circ$ have been removed. The triaxial model is notably overdense
relative to the oblate model in the regions centered on $(\alpha,
\delta)=(125^\circ,50^\circ)$ and $(\alpha,
\delta)=(60^\circ,10^\circ)$. The latter region, located in the
Southern part of the sky, is close to the portion of the Sagittarius
stream excised in our best fit model. The former is coincident with
the Monocerus ring (\citealt{newberg02}), which can be identified, for
example, in Figure 1 of \cite{belokurov06}. We suggest that the
apparent triaxiality may well be caused by the presence of these
overdensities. The increased flexibility of the model can cause it to
`fit' to such substructure and thus the increase in likelihood may be
an artifact.

\begin{figure*}
\centering
\includegraphics[width=15cm,height=3.5cm]{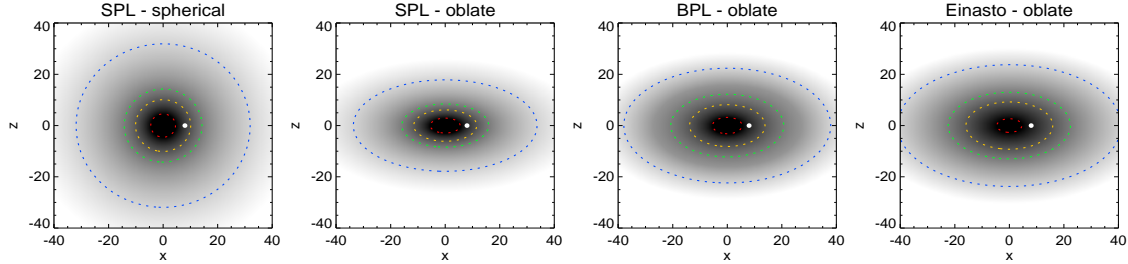}
\caption{\small A side view of our maximum likelihood
  models. Greyscale shows density of stars in plane of Galactocentric
  $(x,z)$ at $y=0$ in four maximum-likelihood models. The blue, green,
  yellow and red contours show density levels corresponding to the
  50th, 90th, 95th and 99th percentiles of the spherical model. The
  white dot marks the location of the Sun.}
\label{fig:models}
\end{figure*}
\begin{figure*}
\centering
    \includegraphics[width=16cm,height=4cm]{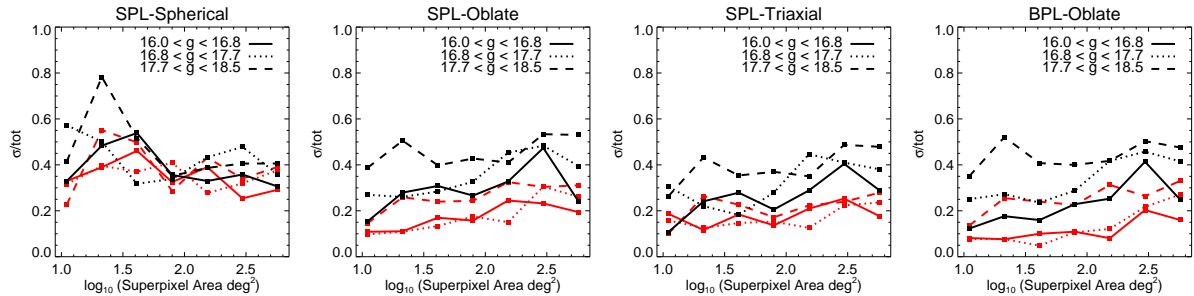}
    \caption{\small The scale dependence of substructure. We show the
      $\sigma/\mathrm{tot}$ fraction as a function of superpixel
      size. Individual panels show the relation for our various
      maximum likelihood models. The different line styles indicate
      different magnitude bins. The black lines show the relation for
      all the stars. The red lines show the relation when stars in the
      vicinity of the Virgo overdensity and Sagittarius have been
      removed.}
\label{fig:superpix}
\end{figure*}

Some earlier investigations have found evidence that the shape of the
stellar halo changes from a flattened distribution at smaller radii to
an almost spherical distribution at larger radii
(e.g. \citealt{preston91}). We can test this claim by allowing the
flattening $q$ to vary with radius. Following the reasoning of
\cite{sluis98}, we make the following substitution
\begin{equation}
  q \rightarrow q\sqrt{\frac{r^2+r^2_0}{q^2r^2+r_0^2}}
\end{equation}
The halo flattening is still $q$ at small radii but tends to
sphericity at large radii. The scale radius $r_0$ determines the
radial range over which this change occurs. For example, for large
values of $r_0$ (e.g. larger than the most distant stars) the
flattening is approximately constant over the applicable radial range.
We fit single power-law models with a varying flattening to the data
and find that large scale radii are preferred ($r_0 > 10^{3}$ kpc)
with an inner flattening of $q=0.53$. This indicates that the
flattening is approximately constant over the radial range of the data
(out to $\sim 40$ kpc). This is in agreement with the deductions of
\cite{sluis98} and \cite{sesar11}, who also found no real evidence for
a varying shape with radius.

In summary, we find that the data are well described by an oblate
density distribution with a constant flattening of $q \sim 0.6$ and a more
steeply declining profile at larger radii. We summarise our maximum
likelihood models by giving a `side view' of the profiles in
Fig. \ref{fig:models}. A spherical model is strongly disfavoured
whereas oblate broken power-law and Einasto models provide a good
representation. 

Our best-fit density distribution can be used to estimate the
  total stellar mass. We find the total number of BHB stars by
  integrating the BHB density profile over all space in the distance range
  $1-40$kpc. The number of BHB stars can be converted
  into a total Luminosity using the relation derived in \cite{deason11} using
  globular clusters:  $N_{\rm BHB}/L \sim 10^{-3}$.  Assuming a
  mass-to-light ratio of $M/L
  \sim 1-5$, the stellar mass is approximately $2-10 \times
  10^{8}M_{\odot}$. \cite{bell08} calculate a total stellar mass of
  $\sim 3.7 \times 10^{8}M_{\odot}$ using main sequence stars, in good
  agreement with our estimate. Note that the total stellar halo mass
  is believed to be $\sim 10^{9}M_{\odot}$ (e.g. \citealt{morrison93}) so we are
  probing a significant fraction of the stellar halo ($\sim 20-100\%$).

\subsection{The Amount of Substructure}

A rough idea of the amount of substructure present in the data can be
attained by computing the rms deviation of the models about the data
($\sigma/\mathrm{tot}$) as defined by equations (2) and (3) in
\cite{bell08}:
\begin{equation}
\langle \sigma^2 \rangle= \frac{1}{n} \sum_k \left( D_k-M_k \right)^2
- \frac{1}{n} \sum_k \left(M^{'}_{k}-M_k \right)^2,
\label{eq:sigma1}
\end{equation}
\begin{equation}
\frac{\sigma}{\mathrm{tot}} ={ \langle \sigma^2 \rangle^{1/2} \over 
 (1/n) \sum_k D_k}.
\label{eq:sigma2}
\end{equation}
Here, $D_k$ is the number of stars in bin $k$, $M_k$ is the expected
number from the model, $M^{'}_{k}$ is a Poisson random deviate from
the model value and $n$ is the total number of bins (in $l$, $b$ and
$m_g$). However, with a total of $\sim 20,000$ stars we can not afford
to use a fixed bin size over the entire sky: pixels must be
simultaneously large enough to contain ample stars and small enough to
adequately sample the substructure. To circumvent this problem, we use
the Voronoi binning method of \cite{cappellari03} to partition the
data in pixels on the sky. This adaptive binning method groups pixels
together, forming `superpixels', with the objective of obtaining a
constant signal-to-noise ratio per bin. We choose the signal-to-noise
ratio of $S/N \sim 4$ (assuming Poisson noise) to ensure the mean
number of stars in each 2D bin is $\gtrsim 10$.  We also check that
our results are not affected by the choice of signal-to-noise. The
stars are split into three $g$ band magnitude ranges, binned into
$\sim 65,000$ $1^\circ \times 1^\circ$ pixels for full sky coverage
which are then combined into $\sim 750$ superpixels using the
procedure described above to estimate the overall
$\sigma/\mathrm{tot}$ values. These are given in the last two columns
of Table \ref{tab:like}. As expected, our $\sigma/\mathrm{tot}$
fraction is significantly reduced when the stars in the region of the
Virgo Overdensity and Sagittarius stream are removed: the estimated
fraction of substructure reduces from $40\%$ to $20\%$. However,
further refinement of our model makes little difference to the
$\sigma/\mathrm{tot}$ fraction, even if the likelihood is
increased. This is indicative of the resilience of our maximum
likelihood method to relatively minor amounts of substructure present
in the data.

We illustrate the dependence of $\sigma/\mathrm{tot}$ on spatial scale
in Fig. \ref{fig:superpix} by grouping together superpixels with
similar spatial scales and computing $\sigma/\mathrm{tot}$ for each
group. This shows how the fraction of substructure depends on the
superpixel size (in $\mathrm{deg}^2$) for different maximum likelihood
models. $\sigma/\mathrm{tot}$ is computed both with (black lines) and
without (red lines) the stars in the vicinity of the Virgo Overdensity
and the Sagittarius stream. The spherical model has relatively large
values of $\sigma/\mathrm{tot}$ over all scales, illustrating the poor
fit of this model to the data. The single and broken power-law oblate
models show a scale dependence, whereby larger scales (larger number
of pixels) have an increased $\sigma/\mathrm{tot}$ fraction relative
to smaller scales. Similar behaviour is also true for the triaxial
model but the $\sigma/\mathrm{tot}$ fractions are slightly lower. As
alluded to earlier, we suspect that this may be caused by the triaxial
model describing some of the large-scale substructure (namely the
Monocerus ring or parts of the Sagittarius stream which have not been
excised).

Different lines in Fig. \ref{fig:superpix} show the dependence of the
$\sigma/\mathrm{tot}$ measure on the apparent magnitude of the tracers
and hence correspond to different distances probed. The slight rise of
the rms with magnitude indicates that higher fractions of substructure
may be present at larger distances. This can be simply explained with
the increase of the substructure lifetime with radius as governed by
the orbital period of the infalling Galactic fragments.
 
When interpreting these plots, it is important to bear in mind that
the smallest superpixels are biased towards the densest parts of the
halo, while the emptiest regions are sampled by the largest
agglomerations of 2D bins. This bias could potentially explain at
least some of the trend with size. On the scales smaller than several
tens of degrees, $\sigma/\mathrm{tot}$ in our models is in the range
$0.05 <\sigma/\mathrm{tot} < 0.2$ indicating that the halo is
\textit{not dominated} by substructure and is relatively smooth. While
there is evidence that the fraction of substructure increases with
scale, this apparent increase in $\sigma/\mathrm{tot}$ could also be
caused by spurious pixels. In sparse regions, the grouping of many
pixels, each containing very few stars, poses a limitation to this
method.


\section{Conclusions}

We have developed a new method to simultaneously model the density
profile of both blue horizontal branch (BHB) and blue straggler (BS)
stars based on their broad-band photometry alone. The probability of
BHB or BS membership is defined by the locus of these stars in $\umg$,
$\gmr$ space. We use these colour-based weights to construct the
overall probability function for the density of the two
populations. When applied to the A-type stars selected from the Sloan
Digital Sky Survey (SDSS) Data Release 8 (DR8) photometric catalog,
the best-fit stellar halo models are identified by applying a maximum
likelihood algorithm to the data.

Based on the data from regions with $30^\circ < |b| < 70^\circ$, we
find that the stellar halo is not spherical in shape, but
flattened. Spherical models cannot reproduce the distribution of the
A-type stars in our sample and are discrepant with the data at both
bright and faint magnitudes. Our best-fit models suggest that the
stellar halo is oblate with a flattening (or minor axis to major axis
ratio) of $q = 0.59$ with a typical uncertainty of $\sigma_q\sim
0.1$. As a simple representation of the stellar halo, we advocate the
use of a broken power-law model with an inner slope $\alphain = 2.3$
and an outer slope $\alphao = 4.6$, the break radius occurring at
about 27 kpc. This gives the formula
\begin{equation}
\rho(\rellip) \propto \begin{cases} r_q^{-2.3} & \rellip \le 27\ {\rm kpc},  \\
                              r_q^{-4.6} & \rellip > 27\ {\rm kpc},
\end{cases}
\end{equation}
with $r_q^2 = R^2 + z^2/0.59^2$. For those who prefer Einasto models,
an equally good law for the stellar halo density is
\begin{equation}
  \rho(\rellip) \propto \exp \left[-4.77[(\rellip/ 20 {\rm
      kpc})^{1/1.7}-1] \right].
\end{equation}
These two formulae are fundamental results of our paper, and can be
summarised as the Milky Way stellar halo is {\it squashed and broken}.
The results are qualitatively similar to those obtained by Sesar et
al. (2011) using main-sequence stars.  There is no evidence in the
data for variation of the flattening with radius.  There is some mild
evidence for a triaxial shape, but the apparent triaxiality might be
due to the presence of the Monocerus ring at low latitudes ($|b| <
40^\circ$) and regions of the Southern Sagittarius stream.

The root-mean-square deviation of the data around the maximum
likelihood model $\sigma/\mathrm{tot}$ typically ranges between $5\%$
and $20\%$. This indicates that the Milky Way stellar halo, or at
least the component traced by the A-type stars in the SDSS DR8, is
{\it smooth} and not dominated by unrelaxed substructure.

This finding is discrepant with the conclusions of
\cite{bell08}. These authors use main sequence turn-off stars selected
from SDSS data release 5 (DR5) to model the stellar halo density. They
argue that no smooth model can describe the data and conclude that the
stellar halo is dominated by substructure (with $\sigma/\mathrm{tot}
\ge 0.33$). We suggest that these contrasting results may be due to
the different methods and tracers used. \cite{bell08} search for the
lowest $\sigma/\mathrm{tot}$ fraction models. However, we find that
the $\sigma/\mathrm{tot}$ values very little between different density
models even if the likelihood is substantially increased. Furthermore,
a lower $\sigma/\mathrm{tot}$ could also indicate that a model is
`fitting' to any substructure (e.g. triaxial models both in this work
and in \citealt{bell08}).  \cite{bell08} also use main sequence stars
as tracers which are more numerous than A-type stars, but are much
poorer distance indicators. The adopted absolute magnitude scale for
main sequence stars is dependent on metallicity and
colour. \cite{bell08} assume a median absolute magnitude of $M_r \sim
4.5$ with a scatter of $\sigma_{M_r} =0.9$; an appropriate choice for
a halo population with metallicity $[\mathrm{Fe/H}] \sim -1.5$. The
scatter, even with an assumed metallicity, is greater than the spread
in absolute magnitude for BS stars ($\sigma_{M_g} \sim 0.5$ per colour
bin). Uncertainties in the distances to stars could lead to compact
substructures becoming more blurred out over a wider range of
distance. It is possible that \cite{bell08} see a somewhat
younger and/or metal-richer component of the stellar halo with the
main sequence tracers. This component could be more unrelaxed than the
one traced by A type stars.

Our conclusion that the stellar halo is composed of a smooth
underlying density, together with some additional substructures such
as the Virgo Overdensity and the Sagittarius Stream, is very
reassuring.  If the stellar halo were merely a hotch-potch of tidal
streams and unrelaxed substructures, then modelling and estimation of
total mass and potential would be much more difficult. Many of the
commonly-used tools of stellar dynamics -- such as the steady-state
Jeans equations -- implicitly assume a well-mixed and smooth
equilibrium. This raises the hope that a full understanding of the
spatial and kinematic properties of stars in the smooth, yet squashed
and broken, stellar halo can yield the gravitational potential and
dark matter profile of the Galaxy itself.

\section*{Acknowledgements}
AJD thanks the Science and Technology Facilities Council (STFC) for
the award of a studentship, whilst VB acknowledges financial support
from the Royal Society. We thank the anonymous referee for many
helpful suggestions.

\label{lastpage}

\end{document}